\documentclass[prd,nofootinbib,superscriptaddress,preprintnumbers,balancelastpage,nofootinbib]{revtex4-2}

\usepackage{graphicx}  
\usepackage{dcolumn}   
\usepackage{bm}        
\usepackage{amssymb,amsmath}
\usepackage{slashed}   
\usepackage[colorlinks=true,allcolors=black]{hyperref}
\usepackage{multirow}
\usepackage{dsfont}
\usepackage{tabularx}
\usepackage[usenames,dvipsnames,svgnames,table]{xcolor}

\def\be{\begin{equation}}
\def\ee{\end{equation}}
\def\bea{\begin{eqnarray}}
\def\eea{\end{eqnarray}}

\newcommand{\Lya}{Ly-$\alpha$ }

\newcommand{\eps}{\epsilon}
\renewcommand{\vec}{\mathbf}

\definecolor{cborange}{HTML}{e69f00}
\definecolor{cbgreen}{HTML}{009e73}
\definecolor{cbyellow}{HTML}{f1dd42}
\definecolor{cblblue}{HTML}{56b4e9}
\definecolor{cbblue}{HTML}{0000FF}
\definecolor{defgrey}{HTML}{808080}
\definecolor{defgreen}{HTML}{008000}
\definecolor{defred}{HTML}{FA5F55}

\def\beq{\begin{equation}}
\def\eeq{\end{equation}}

\def\be{\begin{eqnarray}}
\def\ee{\end{eqnarray}}

\begin{document}
\widetext

\title{Closing in on Pop-III Stars: Constraints and Predictions Across the Spectrum }


\author{Omer Katz}
\affiliation{Raymond and Beverly Sackler School of Physics and Astronomy, Tel-Aviv 69978, Israel}
\author{Diego Redigolo}
\affiliation{INFN, Sezione di Firenze, Via Sansone 1, 50019 Sesto Fiorentino, Italy}
\author{Tomer Volansky}
\affiliation{Raymond and Beverly Sackler School of Physics and Astronomy, Tel-Aviv 69978, Israel}

\begin{abstract}  
The absence of direct high redshift observations poses a significant challenge in understanding the properties of first stars. Nonetheless, the cumulative effect of entire stellar populations can be studied with current data. In this work we use a combination of high redshift observables in order to infer the formation and emission properties of the first stellar populations: high redshift UVLFs, the optical depth of CMB photons to reionization, hydrogen absorption lines in quasar spectra, and measurements of the soft cosmic X-ray background. We study two minimal models of stellar population: i) a single, Pop-II, stellar population which dominates throughout Cosmic Dawn, ii) two distinct stellar populations, Pop-II and Pop-III, dominating at different times with the transition between them taken as a free parameter. We set strong  constraints on the properties of Pop-II stars, and upper limits on the formation and multi-wavelength emission of Pop-III stars. After applying the constraints above, we present the viable envelopes of the 21-cm global signal with and without Pop-III stars. We identify a region in the parameter space of the two population model which predicts a global 21-cm signal distinctive from that of the single population one.  A measurement of such a signal would be a strong indication for the presence of Pop-III stars at early times.  
\end{abstract}

\maketitle
\noindent
\section{Introduction}
While the first stars are too faint to observe individually, the cumulative emission from early stellar populations, spanning a wide range of frequencies, may leave observable signatures in existing data. At its highest  frequencies, which have a large mean free path in both the neutral and ionized universe, X-rays may add a non-negligible contribution to the soft band Cosmic X-ray Background (CXB) as measured today by Chandra~\cite{Fialkov:2016zyq,pochinda2024constrainingpropertiespopulationiii,Lehmer:2012ak,Cappelluti:2012rd}. At lower frequencies, photons in the hard UV range are believed to have driven the epoch of reionization. Indeed, the emission properties of ionizing UV photons and their attenuation in the interstellar medium (ISM) have previously been linked to the CMB optical depth to reionization measured by Planck~\cite{Planck:2018vyg}, hydrogen line absorptions in quasar spectra~\cite{McGreer:2014qwa}, and other reionization observables, as summarized in~\cite{Mason_2019}. Meanwhile, softer UV emission can be used to trace global star formation through high-redshift UV luminosity functions observed by the Hubble Space Telescope (HST)~\cite{Bouwens_2015,Park:2018ljd,Bouwens_2021} and more recently, the James Webb Space Telescope (JWST)~\cite{donnan2024jwstprimernewmultifield}. Lastly, the expected global 21-cm signal, as well as other atomic line signals, encapsulates details of starlight across this entire frequency range~\cite{pochinda2024constrainingpropertiespopulationiii,Munoz:2021psm,Park:2018ljd, Katz:2024ayw, Bernal_2022}. 

The stellar content at Cosmic Dawn can be broadly divided into two distinct populations: (i) the hypothetical first-generation Population III (Pop-III) stars, thought to have formed in pristine, metal-free environments and not yet observed, and (ii) the second-generation Population II (Pop-II) stars, enriched by remnants of the preceding generation. In this work, we combine the above observables, excluding 21-cm data, to set preliminary constraints on the properties of the earliest stellar populations, assuming only minimal modeling~\cite{Munoz:2021psm,Park:2018ljd}. While certain combinations of these observables have been explored in previous studies~\cite{pochinda2024constrainingpropertiespopulationiii, Park:2018ljd,Fialkov:2016zyq,qin2020tale}, this is the first time X-ray, reionization, and UVLFs are jointly applied to study the first stars. Notably, during the preparation of this paper, Ref.~\cite{pochinda2024constrainingpropertiespopulationiii} used multi-wavelength observables to constrain the first stellar populations but excluded UVLFs from their joint likelihood. As already pointed out in Ref.~\cite{Katz:2024ayw}, and further elaborated below in Sec.~\ref{sec:constraints}, UVLFs are essential for constraining the star formation rate density (SFRD). Without UVLFs, the posteriors are either constrained by the priors or dominated by 21-cm data, as in Ref.~\cite{pochinda2024constrainingpropertiespopulationiii}.

The contribution of Pop-III stars to the observed fluxes at different frequencies remains uncertain; they are often considered negligible under the assumption that Pop-III stars are extremely short-lived~\cite{2002A&A...382...28S}, shining primarily at the onset of Cosmic Dawn. In a more bottom-up approach, we analyze two models—one that accounts for Pop-III stars and one that does not. Our findings suggest that a single stellar population suffices to fit the present data, allowing us to establish only upper limits on the properties of Pop-III stars. A promising future observable for probing high redshifts and directly identifying the presence of Pop-III stars is the global 21-cm signal. In Ref.~\cite{Katz:2024ayw}, observations of the Cosmic X-ray Background (CXB), reionization, and the UV Luminosity Functions (UVLFs) were employed to constrain the envelope of global 21-cm signals, assuming only Pop-II stars. Here, we extend this methodology to include a model with Pop-III stars and derive a second envelope. By comparing these envelopes, we identify the range of global 21-cm signals that, within these models, would suggest the presence of an early stellar population.

The results in this work were derived using our new Cosmic Dawn code, which traces the global evolution of star formation, UV and X-ray emission, and the 21-cm differential brightness temperature. It's short running time ($\sim 1$~sec to derive a full global 21-cm evolution) allows us to perform dense and wide scans over the astrophysical parameter space. Despite its simplicity, which allows for the short running time, we are able to reproduce global results derived with the semi analytical 21cmFAST simulation to high precision, as it will be detailed in a forthcoming publication~\cite{Omernew}.

This paper is organized as follows. We begin by deriving the posterior probabilities on the parameters of both stellar models considered here, taking into account all of the above-mentioned observables except for 21-cm data. The modeling is described in the different subsections starting with the SFRD and UVLFs in Sec.~\ref{sec:UVLFandSFRD}, through ionizing UV emission and reionization in Sec.~\ref{sec:CMBUV}, and to X-ray emission and the CXB in Sec.~\ref{sec:Xrays}. The constraint on the \Lya flux is derived in Sec.~\ref{sec:UVLFandLya} as a consequence of the measurements of the UVLF. Our main results are shown in Fig.~\ref{Fig:MCMC}. In Sec.~\ref{sec: The 21cm Brightness Temperature} we present predictions for global 21-cm signals. We show the viable envelope of 21-cm signals for the two stellar population models in Fig.~\ref{fig:21cmsignal} (left), while figure to the right shows in different colors the parameter space of the two stellar population model that cannot be accessed with a single stellar population envelope and leads to a distinguishable 21-cm global signal. In Sec.~\ref{sec:outlook} we conclude. In Appendix~\ref{app:modellingSFRD} we discuss details about the simplified modeling of the SFRD. 

\section{Constraining First Stars}\label{sec:constraints}

In the absence of direct observations during Cosmic Dawn, the individual properties of the first astrophysical objects and of their surroundings remain poorly determined. However models of certain global properties, such as the SFRD, the emission and ISM propagation of ionizing photons,  the ISM attenuated X-ray emissivity and the \Lya emissivity can all be constrained using a combination of observables. In this section we review and update these constraints, applying them to two minimal astrophysical models, based on those implemented in 21cmFAST~\cite{Park:2018ljd,Munoz:2021psm,Mesinger:2010ne,qin2020tale}, and present the full posteriors for these models in Fig.~\ref{Fig:MCMC}. We perform a full scan on the parameters of a model that assumes two stellar populations - one representing the first generation of stars, Pop-III, and the other accounting for the second generation, Pop-II, contaminated by the remnants of the first. Interestingly, current data can already be used to set upper bounds on the formation and emission properties of Pop-III stars. The range of the priors in our scan is shown in Table~\ref{tab:priors}. These are either motivated by high redshift simulations, or chosen to be sufficiently large to capture the full behavior of the likelihood.

\subsection{UVLFs Constraints on the SFRD}\label{sec:UVLFandSFRD}

The UV emission of star forming galaxies is dominated by their massive, short lived, stellar population, making it a common tracer of star formation. Population synthesis models show an essentially constant ratio between the intrinsic luminosity of galaxies at $1500${\AA}, $L_{\rm{UV,}1500}$, and their SFR, $\dot{m}_\star$, both of which  vary from one galaxy to another. This relation can be written as  
\begin{equation}
    \dot{m}_\star = K_{\rm{UV,}1500}\times L_{\rm{UV,}1500} \,,
\label{eq:SFR_prop_L}
\end{equation}
with  $K_{\rm{UV,}1500}=1.15\times 10^{-28}M_\odot\text{yr}^{-1}/\text{ergs } \text{s}^{-1} \text{Hz}^{-1} $ a constant that at leading order is independent of galactic properties. Present studies show that $K_{\rm{UV,}1500}$ varies only by a factor of $\sim2$ across three orders of magnitude in metallicity, hinting for a weak dependence on redshift. Moreover $K_{\rm{UV,}1500}$ has a weak dependence on the initial distribution of stellar masses, at least as long as the distribution is dominated by small mass stars (see Ref.~\cite{Madau_2014} for a detailed review on the subject).

The SFR can be constrained indirectly using the UVLF, $\phi$, which is defined as the comoving galaxy density per absolute magnitude of the galaxy, $M_{\rm UV}$. Below, we will model the SFR in a halo of mass $M_{\rm h}$ as a function of its mass and redshift, which also implies $M_{\rm UV}(z,M_{\rm h})$ through Eq.~\eqref{eq:SFR_prop_L}, and the standard magnitude luminosity relation~\cite{oke1983secondary}. Consequently, we may write the 1500{\AA} UVLF at a given redshift as~\cite{Park:2018ljd}
\begin{equation}\label{eq:UVLF}
    \phi_{1500}  = \frac{dn}{dM_{\rm h}} \left|\frac{dM_{\rm h}}{dM_{{\rm UV},1500}}\right| \,,
\end{equation}
where $dn/dM_{\rm h}$ is the  halo mass function (HMF). Given a measurement of $\phi_{1500}$  one can then determine the SFR for a given choice of HMF. Finally, the SFR density (SFRD) may be obtained using the relation
\begin{equation}\label{eq:SFRD}
    \dot{\rho}_\star(z)=\int_0^\infty \dot{m}_\star(M_h) \frac{dn(M_h,z)}{dM_h}dM_h\,.
\end{equation}

Applying the above prescription and assuming the Sheth-Tormen HMF~\cite{Sheth_1999}, available in the public python toolkit COLOSSUS~\cite{Diemer_2018}, we fit two simplified SFR models. The first, following Ref.~\cite{Park:2018ljd}, assumes a single stellar population (typical considered as Pop-II stars) and has overall four parameters: $t_\star$, $F_\star$, $\alpha_\star$, and $M_{\rm cut}$ such that the SFR is written as
\begin{equation} \label{eq: stellar mass}
    \dot{m}_{\star} = \frac{m_\star}{t_\star H^{-1}(z)}\, \quad ; \quad m_\star(M_h)=M_h\frac{\Omega_{b}}{\Omega_m}f_\star(M_h)\,,
\end{equation}
where the fraction of the baryonic mass within the halo that is in the form of stars is
\begin{equation}\label{eq:SFE powerlaw popII}
    f_\star = f_\star^{(II)}=F_\star^{(II)}\left(\frac{M_h}{10^{10} M_\odot}\right)^{\alpha_\star^{(II)}}e^{-M^{(II)}_{\rm cut}/M_h}\,.
\end{equation}
The power-law behavior is motivated by an array of faint galaxy observations~\cite{behroozi2015simple} while the exponential cutoff is theoretically motivated by inefficient cooling and consequently inefficient star formation in small halos.

\begin{figure*}
\centering
\includegraphics[width=1\textwidth]{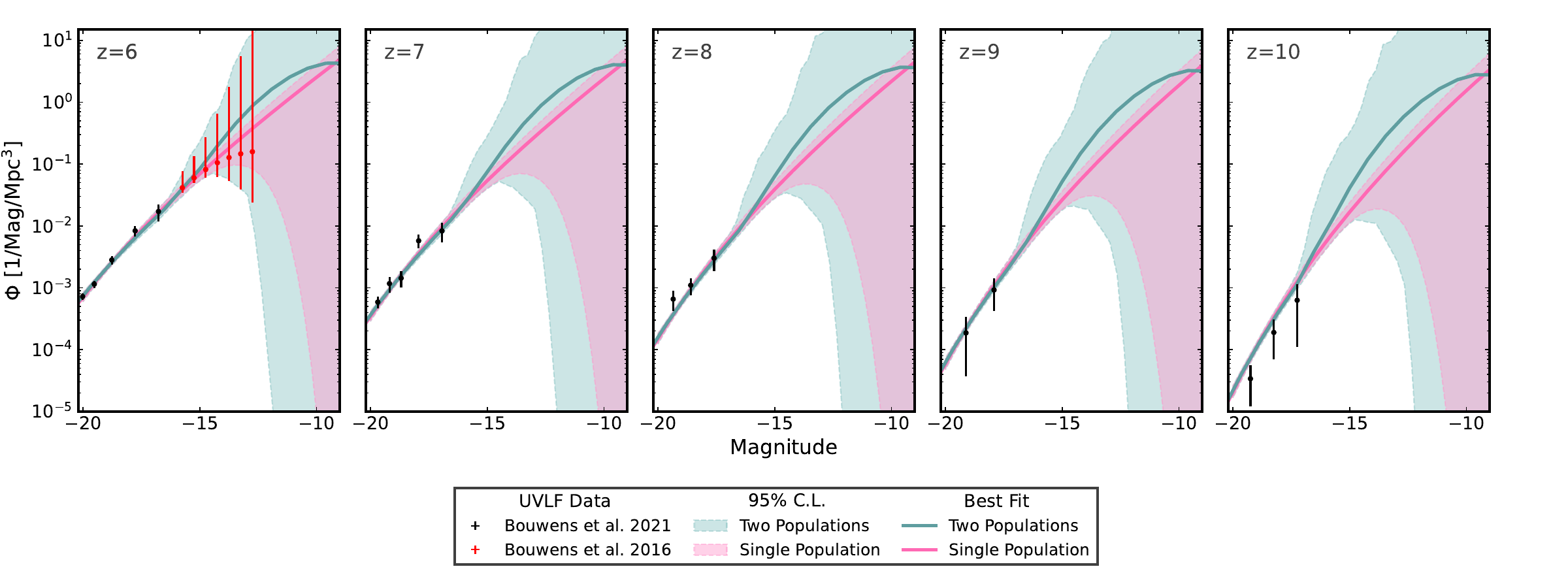}
\qquad
\caption{The $1500${\AA} UVLF as a function of magnitude, assuming a single stellar population (\textbf{pink}), or two stellar populations (\textbf{light blue}). The shaded regions define the 95\% C.L. envelopes, while the solid lines present the best fit models. Each panel correspond to a different redshift. The data used in our study is shown in {\bf black}~\cite{Bouwens_2021} and {\bf red}~\cite{2017ApJ...843..129B}. 
}
\label{Fig:BestFit}
\end{figure*}

In this study, we perform a $\chi^2$-fit of our star formation rate (SFR) model to the Hubble Space Telescope (HST) measurements of the UV luminosity function (UVLF) over the redshift range $z\in[6,10]$. Our analysis incorporates data from several key references \cite{Bouwens:2014fua, 2017ApJ...843..129B, Atek:2018nsc, 2017ApJ...835..113L, 2018ApJ...854...73I, Oesch_2018}, complemented by the recent findings of Bouwens et al. (2021) \cite{Bouwens_2021}. The earlier data were first analyzed in the context of our SFR model by Park et al. \cite{Park:2018ljd}. We focus specifically on fainter galaxies with UV magnitudes $M_{{\rm UV}}>-20$, as these are thought to form in smaller halos with minimal dust attenuation, particularly at high redshift \cite{Yung_2018}. The UVLF for the best-fit parameters of our analysis, along with the corresponding 95\% confidence envelopes, are presented in Fig.~\ref{Fig:BestFit}. We find $\chi^2_{\rm red}=0.64\pm 0.31$  within $1\sigma$ for the best fit model, indicating that no early population is required to explain the present data. Additionally, we assessed the potential impact of incorporating recent data from the James Webb Space Telescope (JWST) \cite{2024MNRAS.533.3222D}, and found that these observations do not significantly affect our best-fit parameters, primarily due to their limited statistical sample. 
However, the inclusion of this data notably worsens the quality of our fit, yielding $\chi_{{\rm red}}^2=1.62\pm0.22$ within $1\sigma$ for the best fit model. If this tension persists in future observations, it would suggest the need for a substantial revision of the current paradigm of star formation models~\cite{Munoz:2024fas}.

To gain insights into the properties of Pop-III stars from current observations, we  perform an additional fit using a star formation rate (SFR) model that incorporates two stellar populations, designated as Pop-II and Pop-III. The SFR for Pop-III stars is modeled similarly to that of Pop-II stars, as described by Eq.~\eqref{eq: stellar mass}. The key difference lies in the baryonic fraction, which has a second cutoff at $M_{\rm cut}^{(II)}$, marking the transition from Pop-III to Pop-II forming halos. This can be expressed as:
\begin{equation}\label{eq:SFE powerlaw popIII}
    f_\star^{(III)}=F_\star^{(III)}\left(\frac{M_h}{10^7 M_\odot}\right)^{\alpha^{(III)}}e^{-M^{(III)}_{\rm cut}/M_h} \ e^{-M_h/M_{\rm cut}^{(II)}}\,.
\end{equation}
We note that both SFRD models share the same $t_*$ and that any possible deviation in $K_{{\rm UV},1500}$ is encapsulated in the normalization parameter $F_\star^{(III)}$.

Pop-III stars are thought to have formed in early, smaller halos. In the absence of metals, cooling in these halos is expected to have occurred primarily through radiative molecular transitions of $H_2$. However, feedback in the form of Lyman-Werner (LW) photons dissociates $H_2$ molecules, inhibiting cooling and ultimately suppressing star formation in the smallest halos.
As structure formation progresses, halos with 
$M_{\rm h}>M_{\rm cut}^{II}$ become abundant. These halos are either large enough to resist feedback or massive enough to cool through atomic hydrogen transitions. By this point, we assume the gas has already been enriched by remnants of the first Pop-III stars and therefore we classify the stars forming in $M_{\rm h}>M_{\rm cut}^{II}$ halos as Pop-II (see Eq.~\eqref{eq:SFE powerlaw popII}). To account for the LW feedback on Pop-III star formation, we adopt a power-law dependence of $M_{\rm cut}^{III}$ on the LW flux, $J_{\rm LW}$, as supported by simulations~\cite{machacek2001simulations,Kulkarni_2021,schauer2021influence}
\begin{equation}\label{eq:LW_feedback}
    M_{\rm cut}^{III} = M_{0}^{III} \times \left(1+A_{\rm LW} \left(\frac{J_{\rm LW}}{10^{-21}\text{erg} \text{ s}^{-1} \text{ cm}^{-2} \text{ Hz}^{-1} \text{ sr}^{-1}}\right)^{\beta_{\rm LW}}\right) \,,
\end{equation}
where  $M_{0}^{III}$
is taken as a free parameter in our analysis, parametrizing additional sources of star formation suppression at small halo mass (for example the relative DM-baryon velocity~\cite{Dalal_2010, Tseliakhovich_2011,Kulkarni_2021,schauer2021influence}). Following~\cite{Munoz:2021psm} we set 
$A_{\rm LW} = 2$ and $\beta_{LW} = 0.6$. These values were chosen to get a functional shape for $M_{\rm cut}^{III}(J_{\rm LW})$ which is in between the two existing simulations of Refs.~\cite{Kulkarni_2021,schauer2021influence}. In appendix~\ref{app:LW} we show that our main results are only mildly sensitive to the specific choice of these parameters within the range set by Refs.~\cite{Kulkarni_2021,schauer2021influence}. Details on the assumed LW emissivity, and the calculation of $J_{\rm LW}$ are given in Sec.~\ref{sec:UVLFandLya} and App.~\ref{app:LW}.

\begin{figure*}%
\centering
\includegraphics[width=1\textwidth]{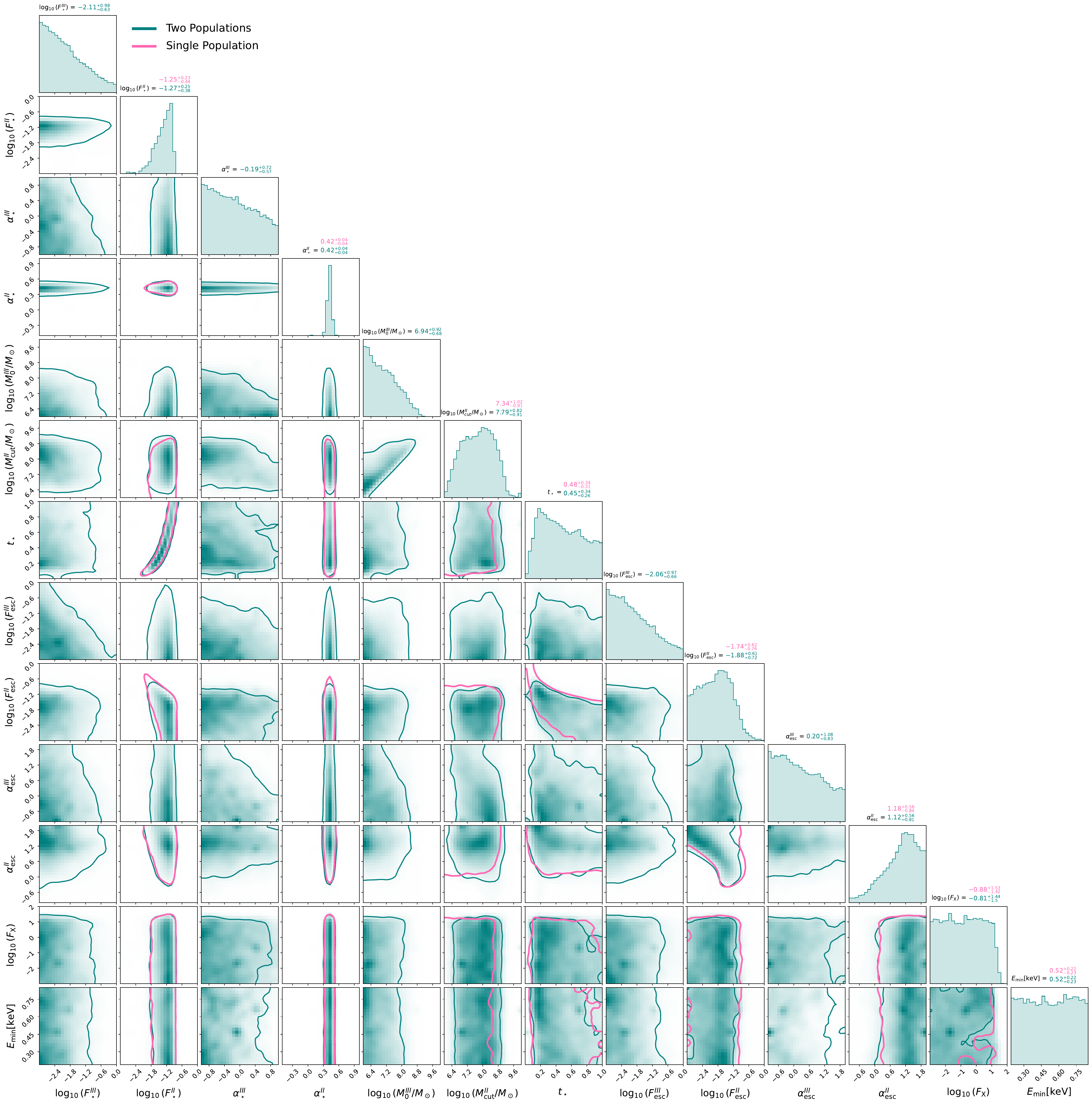}
\qquad
\caption{Corner plot showing the parameter space of the two stellar population model (\textbf{blue}) and single stellar population model (\textbf{pink}), described in Sec.~\ref{sec:UVLFandSFRD}. Upper panels show the 1D PDF of each parameter, with the [16\%, 50\%, 84\%] quantiles displayed above. The best fit parameters are listed in Tab.~\ref{tab:priors}. Contours indicate the $2\sigma$ regions for all parameter pairs, derived by using data of UVLFs measured by the HST~\cite{Bouwens_2015,Bouwens_2021}, $\tau_{\rm e}$ as measured by Planck~\cite{Planck:2018vyg}, \Lya and Ly-$\beta$ absorption lines in quasar spectra~\cite{McGreer:2014qwa}, and the soft CXB measured by Chandra~\cite{Lehmer:2012ak}. Measurements of UVLFs described in Sec.~\ref{sec:UVLFandSFRD} strongly constrain the parameters associated with stellar formation: {$F_\star^i,\alpha_\star^i,M_{\rm cut}^i,t_\star$}, and set upper bounds on those associated with Pop-III.  $\tau_{\rm e}$ together with \Lya and Ly-$\beta$ absorption lines described in Sec.~\ref{sec:CMBUV} constrain the evolution of the neutral hydrogen fraction and thus the parameters associated with stellar formation and the propagation of ionizing photons in the ISM: {$F_{\rm esc}, \alpha_{\rm esc}$}. Measurements of the CXB described in Sec.~\ref{sec:Xrays} set strong upper limits on a the formation of HMXBs, assumed to be the dominant X-ray emitting source at the redshifts of interest, and their X-ray emitting parameters {$F_{\rm X},E_{\rm min}$}.}
\label{Fig:MCMC}
\end{figure*}

Repeating the $\chi^2$-fit for the two stellar population model, we find $\chi^2_{\rm red}=0.72\pm 0.33$ to $1\sigma$,indicating no improvement compared to the previous single population model. This is expected given the absence of higher redshift, lower magnitude data (where lower magnitude corresponds to smaller halos for the range of $\alpha$'s in our scan), where we expect to observe Pop-III stars. However, we may still derive an upper limit on the properties of an early stellar population which we show in Fig.~\ref{Fig:MCMC}. The UVLFs predicted by the best fit models with and without Pop-III together with their 95\% confidence envelopes are also given in Fig.~\ref{Fig:BestFit}.

In the top left panel of Fig.~\ref{fig:global_quantities}, we show the 95\% confidence envelope of the SFRD for both stellar population models presented here. As seen, the two population model can significantly enhance the SFRD even at $z=6$. Specifically, for 8.5\% of models in the the 95\% C.L. the SFRD of Pop-III stars exceeds that of Pop-II~\footnote{Note that this represents the formation rate in terms of mass, not number. While the IMF of Pop-III stars remains uncertain, it is often assumed that the first stars were significantly more massive (see discussion in App.~\ref{app:modellingSFRD}). In this case, the number formation rate of Pop-III stars would be substantially lower than that of Pop-II. }, where for the most extreme model this persists down to $z=5$. While such scenarios would be surprising (see for example Ref.~\cite{Klessen:2023qmc}), we are unaware of any observational evidence that definitively rules them out. Moreover, any upper limit on Pop-III star formation at a given redshift would only serve to tighten constraints on their properties, making our current treatment conservative. Finally, we note that future 21-cm data could place significant constraints on early star formation, as we discuss in  Sec.~\ref{sec:outlook}.

\subsection{CMB and Quasar Spectra Constraints on Ionizing UV Light and Reionization}\label{sec:CMBUV}

The main model-independent constraint on reionization comes from Planck data~\cite{Planck:2018vyg}. Its sensitivity to the evolution of the ionized fraction, $x_{\rm e}=n_e/n_H$, is mostly through the optical depth to reionization which is defined as 
\begin{equation}
\tau_{e}=n_H(z=0) \sigma_T\int^{50}_{0} d z x_e(z)\frac{(1+z)^2}{H(z)}\, ,\label{eq:opticaldepth}
\end{equation}
where $n_H$ is the total number density of hydrogen, $\sigma_T$ is the Thompson cross section, and the somewhat arbitrary integration upper limit is chosen to be high enough to capture the full contribution to $\tau_{e}$ from reionization. The measurements of large scale CMB anisotropies in oupolarization constrain $\tau_{e}= 0.054\pm0.0070$ at $68\%$ C.L.~\cite{Planck:2018vyg}. Additionally, various astrophysical observations suggest that by $z\simeq6$, most of the intergalactic medium (IGM) was already neutral~\cite{Mason_2019}. Throughout this work we use the quoted constraint on $\tau_{\rm e}$, along with the upper limit on the neutral fraction of hydrogen, $x_{HI}<0.06+0.05$ at $z=5.9$, derived from hydrogen absorption lines in quasar spectra~\cite{McGreer:2014qwa}, where the second number indicates the $1\sigma$ uncertainty. 

We consider two ionizing contributions: UV light and X-rays, where in the vast majority of viable models, reionization will be entirely dominated by the UV contribution. Since the mean free path of ionizing UV photons is significantly shorter than that of more energetic X-rays, we model the evolution of the ionized hydrogen fraction, $x_{\rm HII}$, using a two-region approach~\cite{Pritchard:2006sq,Mesinger:2010ne,Park:2018ljd}. At any given redshift, a volume fraction $\chi_{\rm e}(z)$ of the IGM is occupied by ionized bubbles created by UV light from early stars, where hydrogen is assumed to be fully ionized. The remaining regions of the IGM, which UV light has not yet reached, are ionized exclusively by X-ray photons. Overall, the average fraction of ionized hydrogen is  
\begin{equation}
    x_{\rm HII} = \chi_{\rm e} + (1-\chi_{\rm e})\hat{x}_{\rm HII}
\end{equation}
where $\hat{x}_{\rm HII}$ represents the fraction of hydrogen ionized by X-rays in the second region. When calculating the mean ionized fraction, $x_{\rm e}$, we assume that the first helium ionization follows that of hydrogen.

In the UV ionized region, we consider the recombination rate of~\cite{madau1999radiative,Mason_2019}, taking a clumping factor  $C=3$ at reionization as motivated by simulations~\cite{shull2011criticalstarformationratesreionization,Finlator_2012,Kaurov_2015}. The growth rate of the UV ionized volume fraction, $\chi_{\rm e}$, is set by the competition between the aforementioned recombination rate, and the mean ionization rate driven by UV emission from Pop-i stars. The latter is modeled as
\begin{equation}\label{eq:GammaIon}
    \Gamma_{\rm{ion}}^{i} =  \frac{1}{\rho_b^0}\int \frac{d\dot{\rho}^{i}_\star}{dM_{\rm h}} N^{i}_{\rm{ion}}f^{i}_{\rm{esc}}(M_h) dM_{\rm h}\,,
\end{equation}
where $\dot{\rho}_\star^i$ is defined in Eq.~\eqref{eq:SFRD} and $\rho_b^0$ is the energy density of baryons today, $N_{\rm ion}^i$ is the number of ionizing photons emitted per baryon in Pop-II/III stars, and $f_{\rm esc}^i$ is the fraction of ionizing photons that escape to the neutral IGM. We model the escape fraction in the same way as described in Ref.~\cite{Park:2018ljd}, expressing it as 
\begin{equation}
    f^i_{\rm esc}(M_{\rm h}) = F_{\rm esc}^i\left(\frac{M_{\rm h}}{M_{i}}\right)^{\alpha_{\rm esc}^i} \,,
\end{equation}
where $ F_{\rm esc}$ and $\alpha_{\rm esc}$ are free parameters and $M_i=10^{10}M_{\odot},10^7M_{\odot}$ for Pop-II/III respectively.

While simplistic, the homogeneous treatment of UV ionization described above, as implemented in our code, successfully reproduces the results of 21cmFAST simulations~\cite{Park:2018ljd,Mesinger:2010ne} as it will be detailed in a forthcoming publication~\cite{Omernew}. The ionization evolution in the second region, $\hat{x}_{\rm HII}$, is treated as in~\cite{Park:2018ljd, Mesinger:2010ne}, taking the global X-ray emissivity introduced in the following section (see Eq.~\eqref{eq:emissivityshortlife}). 

In Fig.~\ref{Fig:MCMC}, we show the combined constraints from UVLF, reionization, and X-ray observables and in the top right panel of Fig.~\ref{fig:global_quantities} the envelope of the possible redshift evolution of the neutral hydrogen fraction. As we see this envelope is essentially identical for the one-  and the two-population models. This could be attributed to the sensitivity of $\tau_{\rm e}$ to high redshift ionization (see Eq.~\ref{eq:opticaldepth}), which constrains ionization by early Pop-III stars. 
The UVLFs constrain the SFRD as described in Sec.~\ref{sec:UVLFandSFRD}, while the CMB optical depth to reionization and the hydrogen absorption lines in quasar spectra bracket the remaining parameters $F_{\rm esc}^i$, $\alpha_{\rm esc}^i$. We note that the reported values of $F_{\rm esc}^i$ are degenerate with the choices of $N_{\rm ion}^{II}=5000$, $N_{\rm ion}^{III} = 44000$ taken in this work in accordance to population synthesis model~\cite{Leitherer:1999rq,Bromm_2001}.

\subsection{CXB Constraints on X-ray Emission}\label{sec:Xrays}
Measurements of the Cosmic X-ray Background (CXB) in the $[0.5,2]\text{ keV}$ band by Chandra can be used to set upper limits on the X-ray emission of early sources~\cite{Fialkov:2016zyq}. Specifically, here we  focus on the X-ray emission from High Mass X-ray Binaries (HMXBs), which, according to observations of nearby starburst galaxies~\cite{Grimm:2002ta,ranalli20032, Gilfanov:2003bd,Fabbiano:2005pj,Mineo:2011id}, is expected to dominate over alternative sources at high redshifts (with AGNs becoming the primary contribution below $z\sim5$~~\cite{Pacucci:2014wwa}). To remain conservative, we set the upper limit by assuming that all of the CXB measured in the $[0.5,2]\, \text{keV}$ band, which is not attributed to resolved sources, is emitted by HMXBs. We sum  over contributions from all redshifts above $z_{\rm X}^{\rm un}$, below which it is assumed that individual sources are resolved. The 1$\sigma$ constraint, in terms of the global HMXB emissivity $\epsilon_{\rm X}$, is~\footnote{The upper limit in Ref.~\cite{Fialkov:2016zyq} is based on the modeling by Ref.~\cite{Cappelluti:2012rd}. Here we take a more conservative approach, and set the upper limit as $24.3\%$ of the entirety of the soft band CXB measured by Chandra, corresponding to the unresolved contribution.} 
\begin{equation}\label{eq:Chandra}
    \begin{aligned}
        \frac{1}{4\pi}\int_{0.5{\rm keV}}^{2 {\rm keV}} dE E \int_{z_{\rm X}^{\rm un}}^\infty dz \frac{ \epsilon_{\rm X}(E(1+z),z)}{H(z)} e^{-\tau_{\rm X}(E,z_{\rm X}^{\rm un},z)} < 4.06 + 0.29 \frac{\rm keV}  {{\rm cm}^2\text{ sec}\text{ sr}}
    \end{aligned}\,,
\end{equation}
where $\tau_{\rm X}$ is the optical depth of X-ray photons in the IGM~\cite{Mirocha:2014faa}. Because star formation accelerates at lower redshifts, the dominant contribution to the unresolved X-rays comes from the lowest redshift sources. While Chandra resolves soft X-ray galaxies down to $z\sim1$, here we take a conservative choice and assume $z_{\rm X}^{\rm un}=4$. Lowering $z_{\rm X}^{\rm un}$ will significantly strengthen the constraint. 

The short lifetime of HMXBs implies that their luminosity is proportional to their SFR, so that the global emissivity can be written as
\begin{equation}
     \epsilon_X^i(E,t) =\frac{\dot{\rho}^i_\star(t)}{\mu_b}\left\langle\frac{d N_{\rm X}}{dE}\right\rangle\,, \label{eq:emissivityshortlife}
\end{equation}
where $\mu_{\rm b}$ is the mean baryon mass and $\left\langle\frac{d N_{\rm X}}{dE}\right\rangle$ is the average number of X-ray photons emitted per baryon in stars per unit energy. In Appendix~\ref{app:modellingSFRD} we derive the expression for the emissivity above starting from the general definition of global emissivity. In accordance with local measurements~\cite{Grimm:2002ta,ranalli20032, Gilfanov:2003bd,Fabbiano:2005pj,Mineo:2011id} and high redshift simulations~\cite{Fragos:2013bfa,Das:2017fys}, we model the averaged photon spectrum in the $E<15\text{ keV}$ range as a double power-law in energy, matched at $2\text{ keV}$ and truncated at a minimal energy $E_{\rm min}$,
\begin{eqnarray}\label{eq: averaged photon spectrum X}
    \left\langle\frac{d N_X}{dE}\right\rangle &=& \frac{1}{E_0}\left(\frac{E}{E_0}\right)^{-1-\alpha^{\rm s}_X} \Theta \left(E-E_{\mathrm{min}}\right) \Theta \left(2\mathrm{keV}-E\right) +
    \nonumber \\
    &&\frac{1}{E_0}\left(\frac{2\mathrm{keV}}{E_0}\right)^{\alpha^{\rm h}_X-\alpha^{\rm s}_X}\left(\frac{E}{E_0}\right)^{-1-\alpha^{\rm h}_X}\Theta \left(E-2\mathrm{keV}\right) \Theta \left(15\mathrm{keV}-E\right)\,,
\end{eqnarray}
where $E_0$ is a reference energy used for the normalization. The low redshift dominance ensures that we can safely neglect the contribution from hard X-ray emission above $E=15$ keV, and also neglect absorptions in the IGM which are inefficient in an ionized environment, thus removing any possible dependence on reionization through $\tau_{\rm X}$.

The low-energy threshold, $E_{\text{min}}$, is the minimal energy  an X-ray photon must posses in order to escape its host galaxy and penetrate the IGM. Its precise value depends on the assumed local density and metallicity in the galaxy~\cite{Das:2017fys}. For the soft part of the spectrum we take the standard choice of $\alpha^{\rm s}_{\rm X}=1$, whereas for the hard part we take $\alpha^{\rm h}_{\rm X}=2.3$, which is the steepest power-law consistent with population synthesis models, thus leading to the most conservative constraints. 

For an easier comparison with existing literature, we express $E_0$ in terms of the luminoisty-to-SFR ratio over the soft energy band through
\begin{equation}
    \frac{L_{\left[E_\text{min},2\text{keV}\right]}}{\dot{m}_\star} = \frac{1}{\mu_b} \int_{E_{\text{min}}}^{2\text{keV}} E \left\langle \frac{dN_X}{dE}\right\rangle dE \, ,
\end{equation}
where we used $\frac{dL}{dE}=\frac{\dot{m}_\star}{\mu_b} E  \left\langle \frac{dN_X}{dE}\right\rangle$, and apply the common parametrization of $\frac{L_{\left[E_\text{\rm min},2\text{ keV}\right]}}{\dot{m}_\star}=5.55\times10^{-7}F_{\rm X}$, such that the two parameters of the model are $E_{\rm min}$, and $F_{\rm X}$. We will assume for simplicity that $F_{\rm X}$ is a redshift-independent parameter. However, this assumption will need revision to account for the observed anti-correlation between X-ray emissivity and metallicity, as highlighted by Ref.~\cite{Kaur:2022ayn}. This factor could ultimately weaken the constraints derived from our analysis and is postponed to future work~\cite{Xray}\footnote{We thank Andrei Mesinger for many discussions about this point.}. We assume a flat prior for the low-energy cutoff of the soft X-ray band, $E_{{\rm min}}$, within the range $E_\text{min}/{\rm keV}\in\left[0.19-0.85\right]$ which is the $2\sigma$ range where the X-rays optical depth in the ISM, as extracted from the hydrodynamical simulations~\cite{Das:2017fys}, equals one.

\begin{figure}%
\centering
\includegraphics[width=0.4\textwidth]{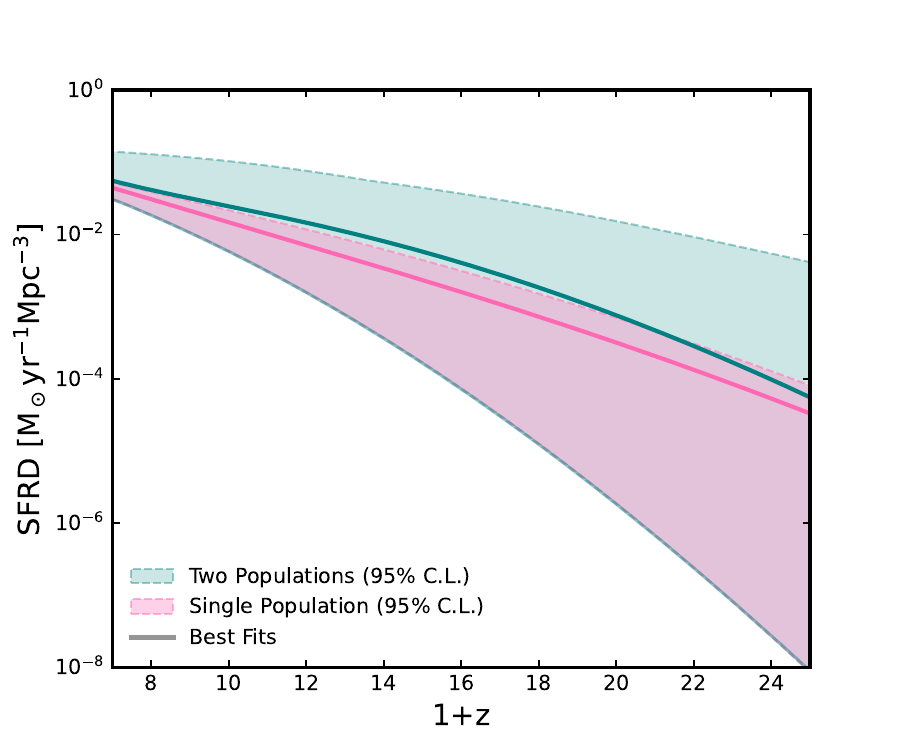}
\includegraphics[width=0.4\textwidth]{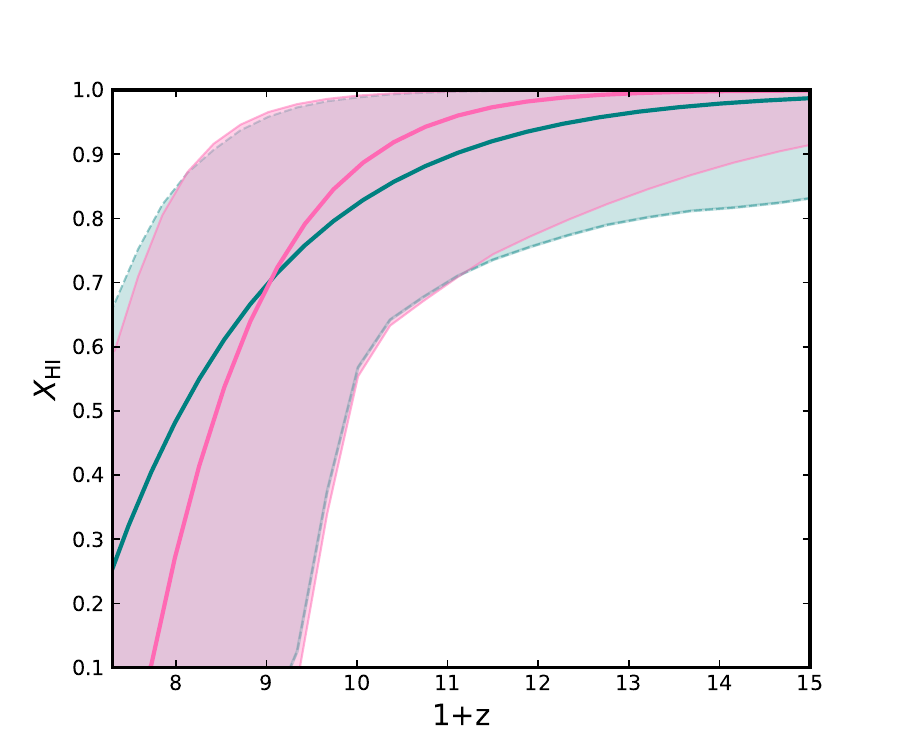}
\includegraphics[width=0.4\textwidth]{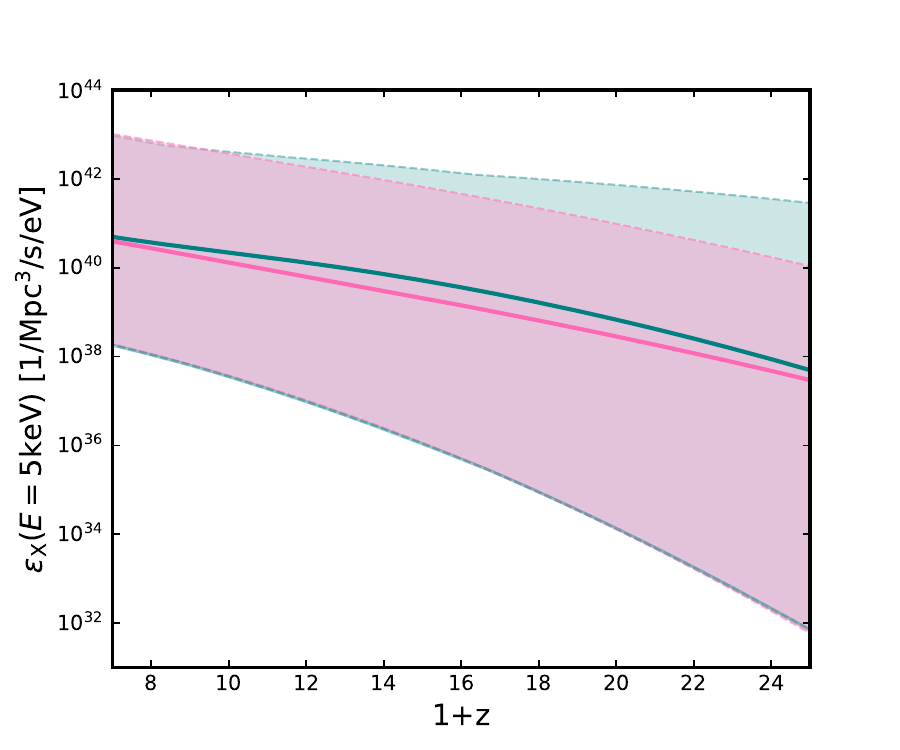}
\includegraphics[width=0.4\textwidth]{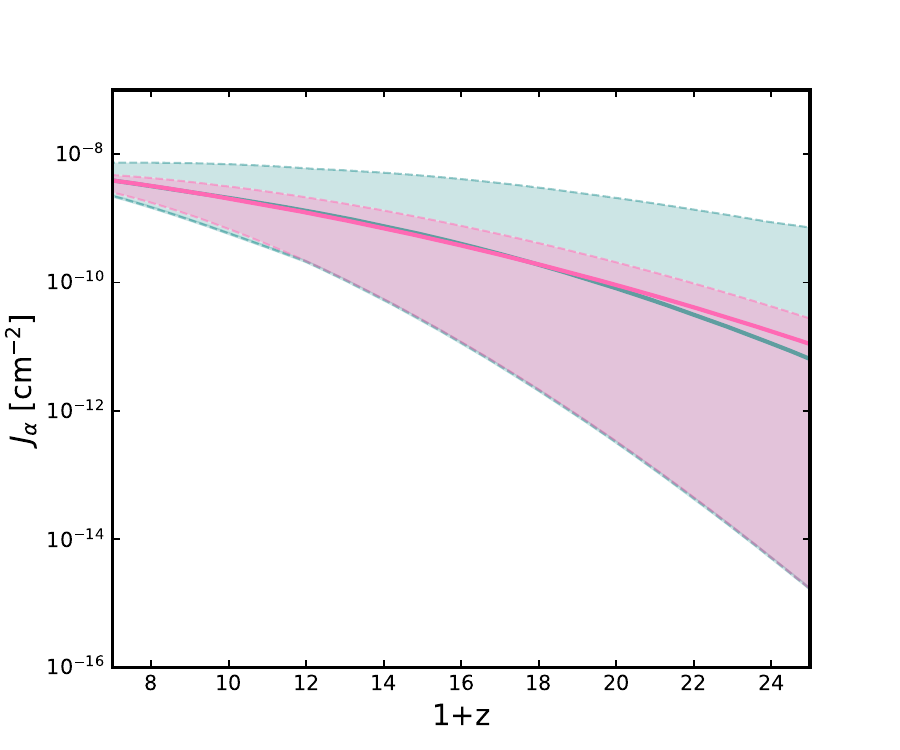}
\qquad
\caption{The evolution of global quantities: the SFRD defined in Eq.~\eqref{eq:SFRD} (\textbf{top left}), neutral fraction of hydrogen (\textbf{top right}), X-ray emissivity defined in Eq.~\eqref{eq:emissivityshortlife} (\textbf{bottom left}) and \Lya flux defined in Eq.~\eqref{sec:UVLFandLya} (\textbf{bottom right}). Results are shown in \textbf{light blue} for the two stellar population model and \textbf{pink} for the single stellar population model. Shaded regions define the 95\% confidence envelope. For each stellar model, we show the evolution across 3000 parameter choices as thin solid lines. The parameters are sampled according to their distribution. The {\bf thick lines} represent the best fit values for the two stellar population models which are summarized in Appendix~\ref{app:modellingSFRD}.}
\label{fig:global_quantities}
\end{figure}

In Fig.~\ref{Fig:MCMC} we show constraints on the two X-ray parameters, $F_X$ and $E_{{\rm min}}$ where the first is bounded from above by Chandra data even though with large uncertainties, while the second is essentially unconstrained apart from its priors set according to simulations. The bound on $F_X$ is  uncorrelated with the other parameters describing stellar formation and propagation of ionizing photon in the ISM and only mildly correlated with $E_{{\rm min}}$. In the bottom left panel of Fig.~\ref{fig:global_quantities} we show the envelope of allowed X-ray emissivities for the two population models introduced in Sec.~\ref{fig:global_quantities} together with their best fit values. From this comparison we see that the two poplulation model allows for a higher X-ray emissivity at high redshift compared to the single population one.

\subsection{UVLF Constraints on the \Lya Emission}\label{sec:UVLFandLya}

The UVLF constraints discussed in Sec.~\ref{sec:UVLFandSFRD} provide a means to constrain the integrated flux of \Lya light, which constitutes a fraction of the UV light emitted. This flux significantly influences the behavior of the 21-cm cosmology signal, as we will elaborate in Sec.~\ref{sec: The 21cm Brightness Temperature}. The constraints on the \Lya flux are illustrated in the bottom right panel of Fig.~\ref{fig:global_quantities} for the two star formation models introduced in Sec.~\ref{sec:UVLFandSFRD}, along with their respective best-fit values. In the remainder of this section, we will connect our modeling of the \Lya flux to the star formation models described in Sec.~\ref{sec:UVLFandSFRD}, enabling us to predict the flux without introducing any additional parameters.

The dominant contribution to the \Lya flux comes from the emission of non-ionizing UV photons, which we parametrize with a global emissivity, $\epsilon_{UV}$. These photons, unable to photoionize the IGM, free-stream until they redshift to the energy of a hydrogen atomic line, where they are immediately absorbed. It is customary to separate the total \Lya flux, $J_\alpha$, into two components: the first, $\bar{J}^{\rm cont}_\alpha$, describes the photons associated with the continuum emission in the [\Lya, Ly-$\beta$] band, which simply redshift to the \Lya frequency. The second component, $\bar{J}^{\rm inj}_\alpha$, represents the \Lya flux arising from cascades of atomic transitions.

Overall, in terms of the UV emissivity, the continuum and injected fluxes are~\cite{Barkana:2004vb,Hirata:2005mz}
\begin{align}
    &\bar{J}_\alpha^{\rm cont}(z) = \frac{(1+z)^2}{4\pi} \int_z^{z_{\max}(2)} \frac{1}{H(z')}\epsilon_{UV}(E_2',z')dz'\,,\label{eq:JalphaCont}\\
    &\bar{J}_\alpha^{\rm inj}(z) \approx\frac{(1+z)^2}{4\pi}\sum_{n=3}^{\infty} P_{n1}\int_z^{z_{\max}(n)} \frac{1}{H(z')}\epsilon_{UV}(E_{n,1}',z')dz'\,, \label{eq:JalphaInt}
\end{align}
where any primed energy $E'_x$ at redhsift $z'$ describes a photon with energy $E_x$ at redhsift $z$: $E'_x=E_x\left(1+z'\right)/\left(1+z \right)$. The maximal redshift is defined as $z_{\max}(n)= z+ \frac{\Delta E_{n}}{E_{n}}(1+z)\, ,$ where $\Delta E_{n}=(1/(n+1)^2-1/n^2)E_{\rm Ry}$ is the energy difference between the $n$ and the $n+1$ energy levels of the hydrogen atom,  and $E_n=(1-1/n^2)E_{\rm Ry}$ with $E_{\rm Ry}\equiv13.6\text{ eV}$ the hydrogen Rydberg energy. Since this splitting goes to zero for large $n$,  the \Lya flux is dominated by the lowest energy levels\footnote{In practice we cutoff the infinite sum in Eq.~\eqref{eq:JalphaInt} at $n=23$, reaching accuracy well below $1\%$ to the injected flux.}.

For the injected flux, stellar emission of a photon initially in the $[n,n+1]$ band with $n \geq 2$ will redshift to the energy of the nearest optically thick $nj$ resonance, where it will excite an electron in the hydrogen atom. The excited hydrogen state will then decay via a cascade until reaching an $n=2$ state. If the $n=2$ state is in an $s$-wave (i.e., $l=0$), the electron can only de-excite to the ground state via two-photon decay. However, if it is in a $p$-wave (i.e., $l=1$), it will transition to the ground state by emitting a \Lya photon. 

Following Ref.~\cite{Hirata:2005mz}, the probability of emitting a \Lya photon through a cascade originating from an optically thick $n,l$ hydrogen excited state can be written recursively as~\footnote{While the density is sufficiently high to keep some forbidden quadrupole excitations optically thick, for the temperatures of interest, the Doppler width of dipole absorption lines is wider than the fine splitting, implying that all excitations are to $l=1$ states~\cite{Hirata:2005mz}.}
\begin{equation}
    P_{n,l} = \frac{\sum_{n'=2}^{n-1} \sum_{l'= 0}^{n'-1} A_{n,l\rightarrow n',l'} P_{n',l'}}{\sum_{n'=2}^{n-1} \sum_{l'= 0}^{n'-1} A_{n,l\rightarrow n',l'}}\ ,
\end{equation}
with the initial values $P_{20}=0$ and $P_{21}=1$, which encode the probabilities for the $2s$ and $2l$ hydrogen levels to emit a \Lya photon. The cascade occurs only through allowed dipole transitions with $\Delta l = \pm 1$, for which the Einstein coefficients $A_{n,l\rightarrow n',l'}$ take a simple analytic form \cite{hoang2005program}. 
Note that direct decays to the $1s$ state result in the emission of a Ly-$n$ photon, which will be immediately reabsorbed by a hydrogen atom and are therefore excluded from the sum.

The emissivity of UV light for a given stellar population, $\epsilon_{\rm UV}^i$, can be written as    
\begin{equation}
\epsilon_{\rm UV} = \frac{\dot{\rho}_\star^i(t)}{\mu_b} \left\langle \frac{d N^i_{\rm UV}}{dE} \right\rangle \,, \label{eq:eL}
\end{equation}
where this formula assumes the short lifetime of the emitting source, and $\left\langle \frac{d N^i_{\rm UV}}{dE} \right\rangle$ is the average number of emitted photons per energy interval by a single baryon in pop-$i$ stars. Since the \Lya band is extremely narrow, the precise spectral shape of $\left\langle \frac{d N^i_{\rm UV}}{dE} \right\rangle$ is of little importance, and $\epsilon_{\rm UV}$ is predominantly set by its normalization. Nevertheless, to remain consistent with previous literature, we adopt the spectral shape over the Lyman band according to population synthesis models~\cite{leitherer1999starburst99,bromm2001generic}, following the procedure outlined in~\cite{Barkana:2004vb}, which is also implemented in 21cmFAST~\cite{Park:2018ljd, Munoz:2021psm, Mesinger:2010ne}. 
In these models, the normalization of the Lyman band spectrum is conveniently set by the total number of ionizing photons discussed in Sec.~\ref{sec:CMBUV}. For completeness, we show the assumed emission spectra in Appendix~\ref{app:modellingSFRD}.

The specific number emissivity derived in this section is also used to calculate the LW flux, which suppresses star formation in small halos, as described in Sec.~\ref{sec:UVLFandSFRD}. For the details of this calculation see App.~\ref{app:LW}.

\section{Sensitivity for Detecting Pop-III Stars with the Global 21-cm Signal}\label{sec: The 21cm Brightness Temperature}

\begin{figure}%
\centering
\includegraphics[width=0.49\textwidth]{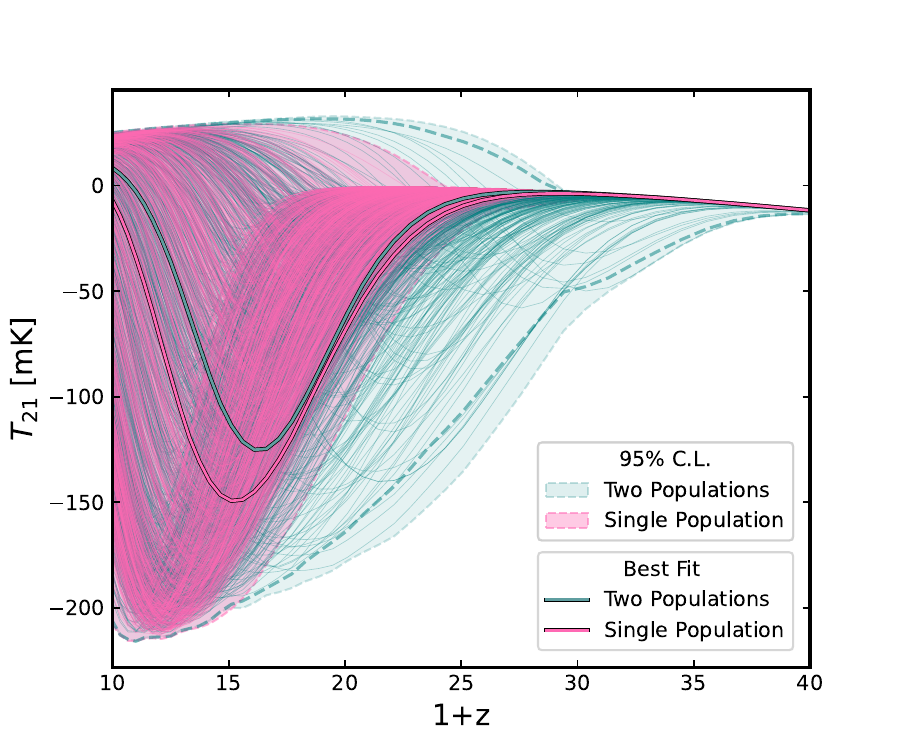}
\includegraphics[width=0.49\textwidth]{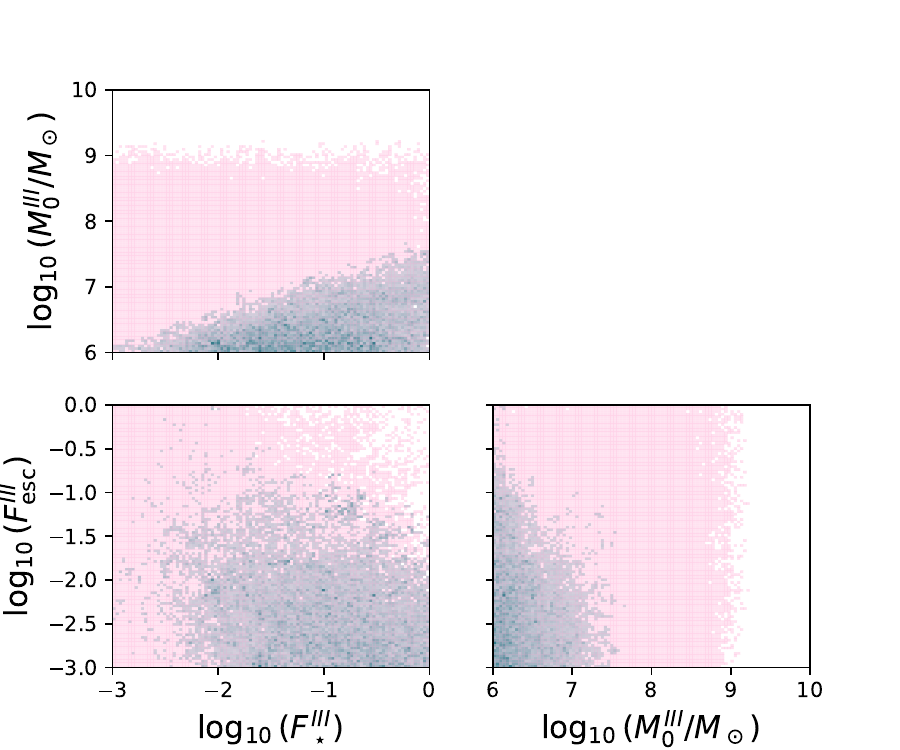}
\qquad
\caption{{\bf Left:} The global 21-cm signals predicted by the viable astrophysical models discussed in Sec.~\ref{sec:constraints}. The {\bf pink} ({\bf light blue}) shaded regions represent the global 21-cm envelopes corresponding to the 95\% C.I.\ of the astrophysical parameters, and thin lines indicate the density of realized models, (using 1000 random parameter choices for each model out of the 2 million generated in the MCMC).  The {\bf thick lines} indicated the predictions for the corresponding best-fit models, as summarized in Tab~\ref{tab:priors}. The {\bf dashed lines} enclose the envelope of models for which the Pop-III star UVLF does not exceed the Pop-II one at $z\leq10$. Current constraints from UVLFs allow for a relatively efficient early Pop-III formation, resulting in a large region of the $T_{21}$ parameter space where the two-population model can be distinguished from the single population one because of its striking global 21-cm signal. 
{\bf Right:} The projections of the two population models are shown on three selected 2D planes. Models corresponding to a 21-cm signal within the 95\% C.I.\ of the single population model (pink-shaded region in the left plot) are highlighted in pink. Models outside this envelope are shaded in blue, with darker regions indicating a higher density of models.}
\label{fig:21cmsignal}
\end{figure}

Hyperfine transitions of ground state hydrogen are accompanied by the absorption and emission of 21-cm photons. These transitions in the IGM are expected to leave a signature over the background radiation parameterized using the brightness temperature (for reviews see~\cite{Pritchard:2011xb,10.1088/2514-3433/ab4a73,Liu:2022iyy}), 
\begin{align}\label{eq:T21}
    T_{21}(z) &= \frac{T_s(z)-T_\gamma(z)}{1+z}\left(1-e^{-\tau_{21}}\right) \,,
\end{align}
where the spin temperature, $T_{\rm s}$, is defined through the relative abundance of triplet states versus singlet states of the hydrogen, $n_1/n_0\equiv 3e^{-E_{21}/T_s}$, and $T_\gamma=2.75(1+z)$K is the 
CMB temperature. The term in brackets accounts for the suppression of a signal which is controlled by the optical depth of the 21-cm photons in the IGM, $\tau_{21}$. The latter can be estimated as $\tau_{21}(z)\simeq 0.03\, x_{HI}(z)\left(\frac{1+z}{25}\right)^{3/2}\left(\frac{10\text{ K}}{T_s}\right)$ and it is directly related to the number density of neutral hydrogen in the IGM, $x_{\rm{HI}}(z)$, which was discussed in relation to UV light from the first stars in Sec.~\ref{sec:CMBUV}.

Currently, the measurement of a global 21-cm signal poses significant challenges   due to the large foreground to signal ratios and possible instrumental systematics. Therefore, despite its high potential in probing early astrophysics, which we  discuss below,  the detailed analysis of global 21-cm measurements is postponed to future work, and we do not include them in the joint constraints of Fig.~\ref{Fig:MCMC}. 
However, looking forward for upcoming improvements, we quantify the sensitivity required to distinguish between the two stellar models considered throughout this work.
 This is demonstrated in Fig.~\ref{fig:21cmsignal} (left) where we show two envelopes of the 21-cm brightness temperature, corresponding to the 95\% C.L. of our astrophysics scans for the two models (Fig.~\ref{Fig:MCMC}). 
The enlarged parameter space of the two population model opens up distinctive feature of the signal which cannot be realized in the single population one, suggesting that future measurements may provide first evidence for the existence of Pop-III stars.

To understand this result we must first dwelve into the strong relation between $T_{\rm s}$, and thus the 21-cm signal, to the astrophysical fluxes we described  in Sec.~\ref{sec:constraints}. This relation was described in detail in Ref.~\cite{Katz:2024ayw}, and it is briefly reviewed here (see also Ref.~\cite{Furlanetto:2006jb} for an original review on the subject). 
The spin temperature is set by the steady state equation
\begin{equation}\label{eq:Ts}
T_s^{-1}=\frac{ T_{\gamma}^{-1}+\bar x_\alpha T_{\alpha}^{-1} +\bar x_k T_{k}^{-1}}{1+\bar x_\alpha+\bar x_k}\,,
\end{equation}
balancing between three distinct spin-flipping processes that control the relative population of singlet and triplet hydrogen states: (i) absorption and emission of CMB photons, which couples $T_{\rm s}$ to the background temperature (ii) neutral hydrogen collisions with itself and with the residual ionized components, coupling $T_{\rm s}$ to the kinetic temperature of the hydrogen in the IGM, and (iii) resonant Raman scattering of hydrogen with \Lya photons produced by the first stars. The last process couples the spin temperature to the effective \Lya color temperature $T_\alpha$, which is tied to $T_{\rm K}$ through the Wouthuysen-Field effect~\cite{1958PIRE...46..240F,1952AJ.....57R..31W}. The interplay between these three different processes is conveniently parametrized by $\bar{x}_\alpha$ and $\bar{x}_{k}$, which are defined as the \Lya and collisional spin-flipping rates normalized by the one of stimulated transitions from CMB photons.

During the epoch of Cosmic Dawn the 21-cm signal is dominated by the astrophysical processes controlled by the light emitted from the first stars. The latter excites, heats, and ionizes neutral hydrogen. Specifically, Lyman-band photons resonantly excite hydrogen, which then decays through a cascade of dipole transitions that eventually end in the production of a \Lya photon, setting the strength of the Wouthuysen-Field effect. Additionally, X-rays photo-ionize hydrogen, depositing heat in the IGM, increasing $T_{\rm K}$. Finally, ionizing UV radiation, which is believed to drive the epoch of reionization, decrease the neutral hydrogen abundance and thus suppress the 21-cm signal regardless of $T_{\rm s}$. 

All in all, the entire dependence of the 21-cm global signal on the astrophysical fluxes can be encapsulated in three quantities which are function of redshift: (i) the global number emissivity of Lyman band photons, $\epsilon_{\rm UV}$, defined in Eq.~\eqref{eq:eL}, (ii) the global number emissivity of X-ray photons, $\epsilon_{\rm X}$, defined in Eq.~\eqref{eq:emissivityshortlife}, and (iii) the global neutral fraction in the IGM, $x_{\rm HI}$, which is mostly sensitive to ionizing UV light. The above  quantities were  modeled and constrained in the previous section, allowing us to 
bracket the envelope of 21-cm signals for a given astrophysical model.

Fig.~\ref{fig:21cmsignal} (left)  shows these envelopes for the two stellar models of Sec.\ref{sec:UVLFandSFRD}. The signals that are consistent with a single stellar populations are shown in pink, while the signals corresponding to two stellar populations are shown in light blue. As expected, the presence of  early Pop-III stars widens the viable envelope. Since the constraining observables are focused at lower redshifts, the upper bound on emission from early stellar populations are rather weak, allowing for a wide range of signals that differ significantly from those of just a single population. On the other hand, since current data do not impose a lower bound on the properties of Pop-III stars, there exists an overlapping region between the two models where the contribution of Pop-III stars is negligible. Overall, we conclude that a deep signal at $z\gtrsim 15$ would be a strong indication of an early second population.
Interestingly, an emission signal at even higher redshifts ($z\gtrsim 20$) remains consistent with present data and would provide strong evidence for an early stellar population with significant X-ray emission. The most extreme features in the left panel of Fig.~\ref{fig:21cmsignal} correspond to models in which the SFR of Pop-III stars dominates over that of Pop-II stars down to low redshift. While these models are not definitively excluded, they are generally disfavored by various astrophysical considerations. For this reason, we indicate with a dashed line the boundary of models where the Pop-II SFR becomes dominant below $z=10$.

A significant and early global 21-cm signal, whether in emission or absorption, requires efficient high-redshift star formation to generate the \Lya photons necessary for the WF coupling. In hierarchical structure formation, this implies that stars must either form in low-mass halos  or have a high star formation efficiency. However, in the single population model sustaining such efficient early star formation would generally lead to a high star formation rate at later times and larger halos, conflicting with existing constraints. In contrast, two-population models predict Pop-III stars to be formed only in small (faint) halos and further incorporate feedback mechanisms that suppress Pop-III star formation at later times, allowing for an early \Lya contribution while remaining consistent with current measurements.  
The key parameters that distinguish a two-population scenario from a single stellar population model are $M_{\rm cut}^{III}$, $F_\star^{III}$, and the escape fraction of ionizing photons, $F_{\rm esc}^{III}$. Fig.~\ref{fig:21cmsignal} (right) illustrates the correlations between these parameters. The pink-shaded regions represent two-population models that yield a 21-cm signal within the 95\% C.I.\ of single-population models (pink region in the left figure), whereas the blue-shaded regions correspond to two-population models that produce a $T_{21}$ signal beyond this envelope.

\section{Summary and Outlook}
\label{sec:outlook}

In this paper, we discuss how combining different observational constraints can provide valuable insights into models of early star formation. Specifically, we use the Hubble measurements of the UV luminosity function to constrain the star formation rate density, CMB anisotropies and quasar spectra to constrain ionizing UV light, and Chandra observations to place an upper bound on the strength of X-ray fluxes. We highlight the significance of these constraints by discussing their impact on global 21-cm data. Our main results are presented in Fig.~\ref{Fig:MCMC} and Fig.~\ref{fig:21cmsignal}, with their key implications summarized as follows:
\begin{itemize}
\item {\bf 
Late Stellar Populations}. For late stellar populations (at $z\lesssim10$), star formation parameters are well determined by current UVLF measurements, except for the minimum halo mass required for star formation. This quantity is only bounded from above, as existing observations primarily probe more massive and luminous halos. Similarly, the star formation rate of early stars is constrained from above, based on their contribution to the low-redshift tail of the signal.
\item {\bf Optical Depth through Reionization}. Unlike other observables, the optical depth through reionization is sensitive to the entire star formation history. Combining the current Planck measurement with the upper limit on the neutral hydrogen fraction at $z\simeq5.9$ from quasar spectra, we can constrain the parameters governing the UV ionizing flux from late stellar populations. However, the contribution from a second, early, stellar populations remains only bounded from above. As expected, the parameters controlling the UV ionizing flux exhibit degeneracies with the normalization of the SFRD.
\item {\bf X-ray Flux Normalization}.  The normalization of the X-ray flux is constrained only from above by Chandra measurements. However, the minimum X-ray energy escaping from galaxies remains largely unconstrained by current data, aside from its anti-correlation with the X-ray flux normalization.

\item {\bf Impact of Early Star Populations on \Lya and 21-cm Signals}. 
Despite the aforementioned upper bounds, an early star population can still produce a significant \Lya flux, potentially altering the expected global 21-cm signal. Consequently, the detection of an absorption feature at $z\gtrsim 15$ or an emission feature at $z\gtrsim 20$ in $T_{21}$ would serve as strong evidence for the existence of an early star population.
\end{itemize}

The constraints derived here will be significantly improved by ongoing and future surveys. The addition of HERA data on high-redshift 21-cm fluctuation~\cite{Abdurashidova_2023} and the assessment of the impact of future data is left for future work. In particular, while current JWST measurements of the UVLFs have not substantially affected our present constraints due to their limited statistics, upcoming data will provide critical insights into the behavior of the UVLF at high redshift. This will, in turn, offer invaluable information on the star formation rate density (SFRD) and its associated UV flux. Additionally, future observations will improve our understanding of ionizing radiation~\cite{Munoz:2024fas}, as advancements in measuring the CMB optical depth to reionization are unavoidably limited by cosmic variance, which restricts any improvement to the current Planck measurement to at most a factor of three~\cite{2012SPIE.8442E..19H,NASAPICO:2019thw}.

Future X-ray telescopes, such as Athena~\cite{Marchesi:2020smf}, are expected to improve the constraints on the unresolved X-ray flux by at least two orders of magnitude, thereby tightening the bound on X-ray emissivity by the same factor. This advancement will greatly enhance our ability to model the high-energy universe and place stronger constraints on the sources of X-ray emission at high redshift.

Beyond their significance for understanding high-redshift stellar evolution, the constraints derived here are crucial for evaluating the potential of 21-cm cosmology to detect deviations from the standard $\Lambda$CDM model. In particular, they help disentangle signals from new physics and uncertainties in astrophysical fluxes. A natural application of this approach is to constrain models of new physics, such as millicharged particles contributing to a small fraction of the dark matter energy density~\cite{Katz:2024ayw}, fuzzy dark matter~\cite{Lazare2024ConstraintsFuzzyDarkMatter}, and dark photons resonantly converting in the interstellar medium~\cite{Pospelov:2018kdh,Caputo:2020avy,Bondarenko:2020moh}, among other scenarios affecting late-Universe dynamics. A detailed assessment of these constraints is left for future work.


\acknowledgments
We are grateful to Andrei Mesinger and Ely Kovets for many useful discussions. OZK thanks the Alexander Zaks scholarship for its support. The work of DR is supported in part by the European Union - Next Generation EU through the PRIN2022 Grant n.~202289JEW4.  The work of DR was performed in part at the Aspen Center for Physics, which is supported by National Science Foundation grant PHY-2210452.
The work of TV is supported, in part, by the Israel Science Foundation (grant No. 1862/21), by the Binational Science Foundation (grant No. 2020220), and by the NSF-BSF (grant No. 2021780).

\bibliographystyle{JHEP}
\bibliography{bib.bib}

\clearpage
\newpage
\appendix

\section{Constraining Procedure}\label{sec:Likelihood}

\begin{table*}[]
\renewcommand{\arraystretch}{1.5}
\begin{tabular}{ccccclcclllll}
\multicolumn{8}{c}{Star Formation}                                                                                                                                                                                                                                                                                                                                                                                                                                                                                       &  &  &  &  &  \\ \cline{1-8}
\multicolumn{1}{|l|}{Population} & \multicolumn{3}{c|}{Pop-III}                                                                                                                                                & \multicolumn{3}{c|}{Pop-II}                                                                                                                                                                                                                           & \multicolumn{1}{c|}{-}                          &  &  &  &  &  \\ \cline{1-8}
\multicolumn{1}{|l|}{Parameter}  & \multicolumn{1}{c|}{$\log_{10}(F_{\star}^{\rm III})$}      & \multicolumn{1}{c|}{$\alpha_{\star}^{\rm III}$}   & \multicolumn{1}{c|}{$\log_{10}(M_{0}^{\rm III})$ $\left[ M_\odot \right]$} & \multicolumn{1}{c|}{$\log_{10}(F_{\star}^{\rm II})$}        & \multicolumn{1}{c|}{$\alpha_{\star}^{\rm II}$} & \multicolumn{1}{c|}{$\log_{10}(M_{\rm cut}^{\rm II})$ $\left[ M_\odot \right]$}                                                                              & \multicolumn{1}{c|}{$t_\star$}                  &  &  &  &  &  \\ \cline{1-8}
\multicolumn{1}{|c|}{Priors}     & \multicolumn{1}{c|}{$\left[ -3,0 \right]$} & \multicolumn{1}{c|}{$\left[ -1,1 \right]$}        & \multicolumn{1}{c|}{$\left[ 6,10 \right]$}                  & \multicolumn{1}{c|}{$\left[ -3,0 \right]$}  & \multicolumn{1}{c|}{$\left[ -0.5,1 \right]$}     & \multicolumn{1}{c|}{\begin{tabular}[c]{@{}c@{}}$\left[ 6,10 \right]$\\ and\\ $M_{\rm cut}^{\rm II}>M_{0}^{\rm III}$\end{tabular}} & \multicolumn{1}{c|}{$\left[ 0,1 \right]$}       &  &  &  &  &  \\ \cline{1-8}
\multicolumn{1}{|c|}{Best fit}   
& \multicolumn{1}{c|}{\begin{tabular}[c]{@{}c@{}}\textcolor{teal}{-1.70} \\ \end{tabular}} 
& \multicolumn{1}{c|}{{\begin{tabular}[c]{@{}c@{}}\textcolor{teal}{0.93} \\ \end{tabular}}} 
& \multicolumn{1}{c|}{\begin{tabular}[c]{@{}c@{}}\textcolor{teal}{6.70} \\ \end{tabular}}     
& \multicolumn{1}{c|}{\begin{tabular}[c]{@{}c@{}}\textcolor{teal}{-1.07} \\ 
\textcolor{purple}{-2.07}\end{tabular}}          
& \multicolumn{1}{c|}{\begin{tabular}[c]{@{}c@{}}\textcolor{teal}{0.48} \\ 
\textcolor{purple}{0.42}\end{tabular}} 
& \multicolumn{1}{c|}{\begin{tabular}[c]{@{}c@{}}\textcolor{teal}{8.15} \\ 
\textcolor{purple}{6.05}\end{tabular}} 
& \multicolumn{1}{c|}{\begin{tabular}[c]{@{}c@{}}\textcolor{teal}{0.77}\\ 
\textcolor{purple}{0.07}\end{tabular}}                                                                  &  &  &  &  &  \\ \cline{1-8}
\multicolumn{1}{|c|}{68\% C.I.}
& \multicolumn{1}{c|}{\begin{tabular}[c]{@{}c@{}} \textcolor{teal}{$-2.11^{+0.98}_{-0.63}$} \\ \end{tabular}} 
& \multicolumn{1}{c|}{\begin{tabular}[c]{@{}c@{}} \textcolor{teal}{$-0.19^{+0.72}_{-0.57}$} \\ \end{tabular}} 
& \multicolumn{1}{c|}{\begin{tabular}[c]{@{}c@{}} \textcolor{teal}{$6.94^{+0.92}_{-0.68}$} \\ \end{tabular}} 
& \multicolumn{1}{c|}{\begin{tabular}[c]{@{}c@{}} \textcolor{teal}{$-1.27^{+0.25}_{-0.38}$} \\ \textcolor{purple}{$-1.25^{+0.23}_{-0.44}$}\end{tabular}} 
& \multicolumn{1}{c|}{\begin{tabular}[c]{@{}c@{}} \textcolor{teal}{$0.42^{+0.04}_{-0.04}$} \\ \textcolor{purple}{$0.42^{+0.04}_{-0.04}$}\end{tabular}} 
& \multicolumn{1}{c|}{\begin{tabular}[c]{@{}c@{}} \textcolor{teal}{$7.79^{+0.82}_{-0.91}$} \\ \textcolor{purple}{$7.34^{+1.02}_{-0.91}$}\end{tabular}} 
& \multicolumn{1}{c|}{\begin{tabular}[c]{@{}c@{}} \textcolor{teal}{$0.45^{+0.34}_{-0.26}$}\\ \textcolor{purple}{$0.48^{+0.34}_{-0.31}$}
\\ \end{tabular}} 
                         &  &  &  &  &  \\ \cline{1-8}
\multicolumn{1}{l}{}             & \multicolumn{1}{l}{}                            & \multicolumn{1}{l}{}                              & \multicolumn{1}{l}{}                                                  & \multicolumn{1}{l}{}                             &                                                & \multicolumn{1}{l}{}                                                                                                                              & \multicolumn{1}{l}{}                            &  &  &  &  &  \\
\multicolumn{1}{l}{}             & \multicolumn{4}{c}{Reionization}                                                                                                                                                                                               &                                                & \multicolumn{2}{c}{X-rays}                                                                                                                                                                          &  &  &  &  &  \\ \cline{1-5} \cline{7-8}
\multicolumn{1}{|c|}{Population} & \multicolumn{2}{c|}{Pop-III}                                                                        & \multicolumn{2}{c|}{Pop-II}                                                                                              & \multicolumn{1}{l|}{}                          & \multicolumn{2}{c|}{-}                                                                                                                                                                              &  &  &  &  &  \\ \cline{1-5} \cline{7-8}
\multicolumn{1}{|c|}{Parameter}  & \multicolumn{1}{c|}{$\log_{10}(F_{\rm esc}^{\rm III})$}    & \multicolumn{1}{c|}{$\alpha_{\rm esc}^{\rm III}$} & \multicolumn{1}{c|}{$\log_{10}(F_{\rm esc}^{\rm II})$}                           & \multicolumn{1}{c|}{$\alpha_{\rm esc}^{\rm II}$} & \multicolumn{1}{l|}{}                          & \multicolumn{1}{c|}{$\log_{10}(F_{\rm X})$}                                                                                                                  & \multicolumn{1}{c|}{$E_{\rm min} [{\rm keV}]$}  &  &  &  &  &  \\ \cline{1-5} \cline{7-8}
\multicolumn{1}{|c|}{Priors}     & \multicolumn{1}{c|}{$\left[ -3,0 \right]$} & \multicolumn{1}{c|}{$\left[ -1,2 \right]$}        & \multicolumn{1}{c|}{$\left[ -3,0 \right]$}                       & \multicolumn{1}{c|}{$\left[ -1,2 \right]$}       & \multicolumn{1}{l|}{}                          & \multicolumn{1}{c|}{$\left[ -3,2 \right]$}                                                                                              & \multicolumn{1}{c|}{$\left[ 0.19,0.85 \right]$} &  &  &  &  &  \\ \cline{1-5} \cline{7-8}
\multicolumn{1}{|c|}{Best fit}   
& \multicolumn{1}{c|}{\begin{tabular}[c]{@{}c@{}}\textcolor{teal}{-2.73}\\ \end{tabular}}    
& \multicolumn{1}{c|}{\begin{tabular}[c]{@{}c@{}}\textcolor{teal}{-0.84}\\ \end{tabular}}    
& \multicolumn{1}{c|}{\begin{tabular}[c]{@{}c@{}}\textcolor{teal}{-1.49}\\ \textcolor{purple}{-0.58}\end{tabular}}    
& \multicolumn{1}{c|}{\begin{tabular}[c]{@{}c@{}}\textcolor{teal}{-0.11}\\ \textcolor{purple}{1.82}\end{tabular}}    
& \multicolumn{1}{l|}{}                           
& \multicolumn{1}{c|}{-}                                 & \multicolumn{1}{c|}{-}                          

                        &  &  &  &  &  \\ \cline{1-5} \cline{7-8}

\multicolumn{1}{|c|}{68\% C.I.}  
& \multicolumn{1}{c|}{\begin{tabular}[c]{@{}c@{}} \textcolor{teal}{$-2.06^{+0.97}_{-0.66}$} \\ \end{tabular}} 
& \multicolumn{1}{c|}{\begin{tabular}[c]{@{}c@{}} \textcolor{teal}{$0.20^{+1.08}_{-0.83}$} \\ \end{tabular}} 
& \multicolumn{1}{c|}{\begin{tabular}[c]{@{}c@{}} \textcolor{teal}{$-1.88^{+0.61}_{-0.72}$} \\ \textcolor{purple}{$-1.74^{+0.62}_{-0.76}$}\end{tabular}} 
& \multicolumn{1}{c|}{\begin{tabular}[c]{@{}c@{}} \textcolor{teal}{$1.12^{+0.56}_{-0.81}$} \\ \textcolor{purple}{$1.18^{+0.56}_{-0.84}$}\end{tabular}} 
& \multicolumn{1}{l|}{} 
& \multicolumn{1}{c|}{\begin{tabular}[c]{@{}c@{}} \textcolor{teal}{$-0.81^{+1.44}_{-1.50}$} \\ \textcolor{purple}{$-0.88^{+1.53}_{-1.42}$}\end{tabular}} 
& \multicolumn{1}{c|}{\begin{tabular}[c]{@{}c@{}} \textcolor{teal}{$0.52^{+0.22}_{-0.23}$} \\ \textcolor{purple}{$0.52^{+0.22}_{-0.23}$}\end{tabular}} 

&  &  &  &  &  \\ \cline{1-5} \cline{7-8}
\end{tabular}
\caption{The parameter range, highest likelihood values and the 68\% C.I. for the two stellar population model (black/teal) and the single stellar population model (black/pink) presented in Sec.~\ref{sec:UVLFandSFRD}.  The parameters associated with star formation, introduced in Sec.~\ref{sec:UVLFandSFRD}, are shown in the {\bf top table}, the ones controlling  reionization properties as discussed in Sec.~\ref{sec:CMBUV} are displayed on the {\bf bottom left}, while those controlling the X-ray properties considered in Sec.~\ref{sec:Xrays} are on the {\bf bottom right}. Parameters that vary by more than an order of magnitude are sampled in log-space. The highest likelihood values are acquired in our analysis according to~\eqref{eq:likelihood}. Given the flatness of the X-ray contribution in the likelihood, there is a degeneracy in the most likely X-ray parameters which is only  broken by the effect of X-rays on reionization, which remains negligible over the majority of the parameter space. We therefore take $F_{\rm X}=0.14$ and $E_{\rm min} = 0.5$ to match the median values (see Fig.~\ref{Fig:MCMC}) whenever referring to the best fit X-ray values.}\label{tab:priors}
\end{table*}

To produce Fig.~\ref{Fig:MCMC} we sampled our joint likelihood function, $\mathcal{L}$, using the public MCMC sampler emcee~\cite{Foreman_Mackey_2013}. The joint likelihood function was constructed as the product 
\begin{equation}\label{eq:likelihood}
\mathcal{L} = \mathcal{L}_{\rm UVLF} \times \mathcal{L}_{\tau_{\rm e}} \times \mathcal{L}_{Q} \times \mathcal{L}_{CXB} \,, 
\end{equation} 
where the likelihood functions on the right-hand side correspond to the high redshift observables discussed in Sec.~\ref{sec:constraints}. From left to right, these are measurements of high redshift UVLFs, the optical depth to reionization, hydrogen absorption lines in quasar spectra, and the CXB. 

At each observed redshift, the UVLF likelihood is taken as a Gaussian such that
\begin{equation}
    \log\left( \mathcal{L}^{\rm z}_{\rm UVLF} \right) = -\sum_i\frac{\left(\phi_{\rm obs}^i(M_{\rm UV}^i)-\phi_{\rm model}^i \left(M_{\rm UV}^i;\ \vec{\theta}\right)\right)^2}{\sigma_{i,+/-}^2} \,,
\end{equation}\label{eq:UVLF_likelihood}
where the index $i$ runs over all data points with magnitude $M_{\rm UV} > -20$ collected at redshift $z$, and tabulated in~\cite{Bouwens_2021}. $\phi^i_{\rm obs}$ and $\phi^i_{\rm model}$ represent the observed and modeled UVLFs, and $\vec{\theta}$ denotes the astrophysical parameters. Since some data points have asymmetric errors, we define $\sigma_{i,\pm}^2$ as the upper $1\sigma$ value when $\phi^i_{\rm model} > \phi^i_{\rm obs}$ and the lower $1\sigma$ value when $\phi^i_{\rm model} < \phi^i_{\rm obs}$. The UVLF likelihood in Eq.~\eqref{eq:likelihood} is calculated by multiplying $\mathcal{L}^{\rm z}_{\rm UVLF}$ across all observed redshifts from $z = 6$ and above.

To evaluate the likelihoods functions associated with reionization observables we model the full evolution of $x_{\rm HI}(z)$ and calculate $\tau_{\rm e}$ (see Eq.~\eqref{eq:opticaldepth}) at each call. In accordance with quasar spectra observations, which provide a $1\sigma$ upper limit of $x_{\rm HI} < 0.05 + 0.06$, and Planck's measurements that imply $\tau_{\rm e} = 0.054 \pm 0.007$ at $68\%$ C.L., we define the following functions:
\begin{equation}
    \log\left( \mathcal{L}_{\rm \tau_{\rm e}} \right) = -\frac{\left(0.054-\tau_{\rm e}(\vec{\theta})\right)^2}{0.007^2} \,,
\end{equation}
and
\begin{equation}
    \log\left( \mathcal{L}_{\rm Q} \right) =
    \begin{cases}
     -\frac{\left(0.05-x_{\rm HI}(z=5.9;\vec{\theta})\right)^2}{0.06^2} &\text{if } x_{\rm HI}(z=5.9;\vec{\theta})>0.05 \\
    0 & \text{else}
\end{cases}
\\.
\end{equation}
Additionally, in each likelihood call we also model the contribution to the CXB in the $[0.5,2]$ keV band and evaluate
\begin{equation}
    \log\left( \mathcal{L}_{\rm CXB} \right) =
    \begin{cases}
     -\frac{\left(4.06-I^{\rm X}_{[0.5,2]}(z_{\rm un}^{\rm X}=4;\vec{\theta})\right)^2}{0.29^2} &\text{if } I^{\rm X}_{[0.5,2]}(z_{\rm un}^{\rm X}=4;\vec{\theta})>4.06\frac{\rm keV}{{\rm cm}^2\text{ sec}\text{ sr}} \\
    0 & \text{else}
\end{cases}
\\,
\end{equation}
in accordance to the $I^{\rm X}_{[0.5,2]}=4.06\pm0.29\frac{\rm keV}{{\rm cm}^2 \text{ sec}\text{ sr}}$ Chandra constraint. As already discussed,  we treat the Chandra result only as an upper limit on the X-ray emission of HMXBs given the unknown contribution from additional sources.

The priors for the scan are listed in Tab.~\ref{tab:priors}. The range for these priors is chosen to be sufficiently large to capture the behavior of the corresponding PDFs. Parameters that vary over more than an order of magnitude are sampled in log-space.

\section{Emissivity modeling}\label{app:modellingSFRD}

Here we delve into the approximation that lead to the emissivity formulas in Eq.~\eqref{eq:emissivityshortlife} and Eq.~\eqref{eq:eL}. We begin  by writing a general equation for the volume averaged comoving number emissivity, $\epsilon^i$, (henceforth emissivity) of a stellar population $i$.
For this, we define $dL^i(E,m,t)/dE$ as the specific luminosity of a single population $i$ star, with mass $m$, at age $t$. The emissivity of the entire population is given by summing up the contributions from all active pop-$i$ stars in a differential comoving volume
\begin{equation} \label{eq: emissivityApp}
    \epsilon_i(E,t) = \int_0^{\infty} \int_{t_\star}^t \frac{dL^i(E,m'_*,t'-t_\star)}{EdE}\frac{\dot{\rho}_\star^i(t+t_\star -t')\mathcal{F}_i(m'_*)}{m_*'}dt'dm_*'\,,
\end{equation}
where $t_\star$ is the time at the beginning of the Cosmic Dawn, $\dot{\rho}^i_\star$ is the pop-$i$ SFRD as usual, and the factor $1/E$ is essential to obtain an emissivity expressed in terms of photon number rather than energy. Finally, $\mathcal{F}_i(m)$ is the PDF of pop-$i$ stellar masses and can be related to the Initial Mass Function (IMF), $\xi^i(m)$, by
\begin{equation}
\mathcal{F}_i(m) dm \equiv \frac{\xi^i(m)mdm}{\int \xi^i(m)mdm}\ .\label{eq:IMF}
\end{equation}

Interestingly, if the lifetime of the luminous objects is very short compared to the typical timescale of variations in the SFRD (which is roughly of the order of the cosmological horizon at a given reshift) then the number density of photons can be considered as a delta function in time 
\begin{equation}\label{eq:deltaSpectrum}
\frac{dL^i_*(E,m_*,t)}{EdE}=\frac{d N^i_*}{dE}(E,m_*)\delta(t)\, .
\end{equation}
In this limit Eq.~\eqref{eq: emissivityApp} simplifies to  
\begin{equation}
     \epsilon_i(E,t) =\frac{\dot{\rho}_\star^i(t)}{\mu_b}\left\langle\frac{d N^i_*}{dE}\right\rangle\,. \label{eq:emissivityshortlifeApp}
\end{equation}
which is the equation for the emissivity used in the main text. In its simplified form, the time dependence of the emissivity is entirely encoded in the SFRD while the energy dependence is encoded in the averaged photon spectrum, weighted with the PDF of the luminous object population,
\begin{equation}
\left\langle\frac{d N_*^i}{dE}\right\rangle\equiv \int_0^\infty \frac{d N_*^i}{dE}(E,m_*')\mathcal{F}_i(m'_*)\frac{\mu_b dm_*'}{m_*'}. \label{eq:average}
\end{equation}
We remind the reader that $\left\langle\frac{d N_*^i}{dE}\right\rangle$ has the meaning of the average number of emitted photons per energy interval emitted by a single baryon in pop-$i$ stars. On the left of  Fig.~\ref{fig:SFRD_timescale} we show for completeness the emission spectra used in this work which are derived from population synthesis models~\cite{leitherer1999starburst99,bromm2001generic}, following the procedure outlined in~\cite{Barkana:2004vb}, which is implemented in 21cmFAST~\cite{Park:2018ljd, Munoz:2021psm, Mesinger:2010ne}. 

The lifetimes of observed high-mass X-ray binaries and UV-emitting Population II stars are relatively short, which justifies using the simplified expression in Eq.~\eqref{eq:emissivityshortlifeApp} to calculate the X-ray emissivity from HMXBs and the UV emissivity from Pop-II stars. In the following, we explore the conditions under which this approximation is valid for Pop-III stars.

\subsection{Pop-III star Lifetime}

\begin{figure}[t] 
  \centering
  \includegraphics[width=0.49\textwidth]{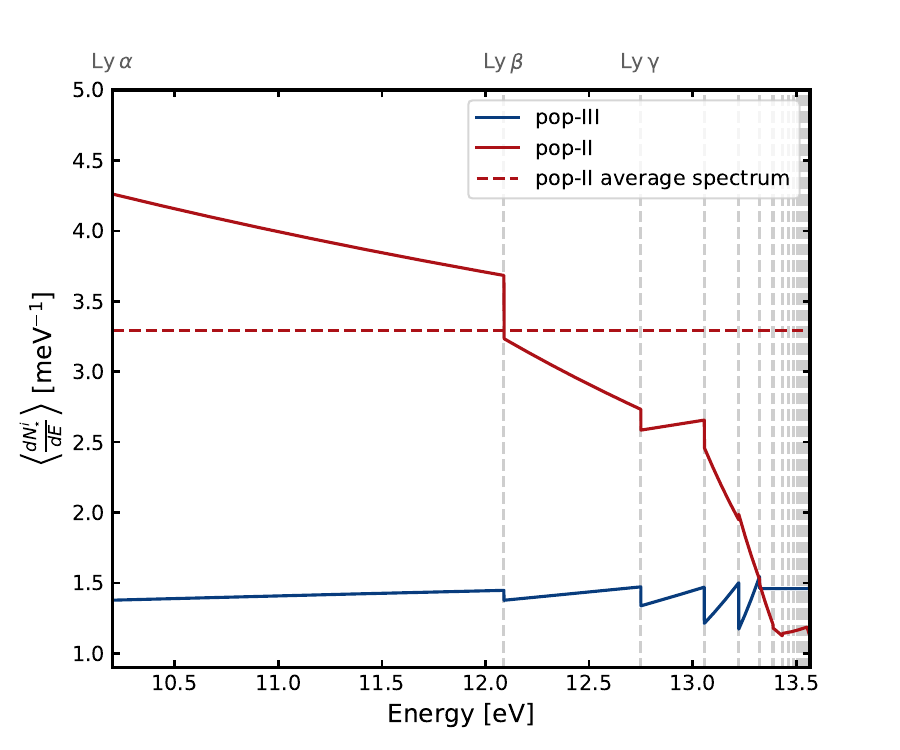}
  \includegraphics[width=0.49\textwidth]{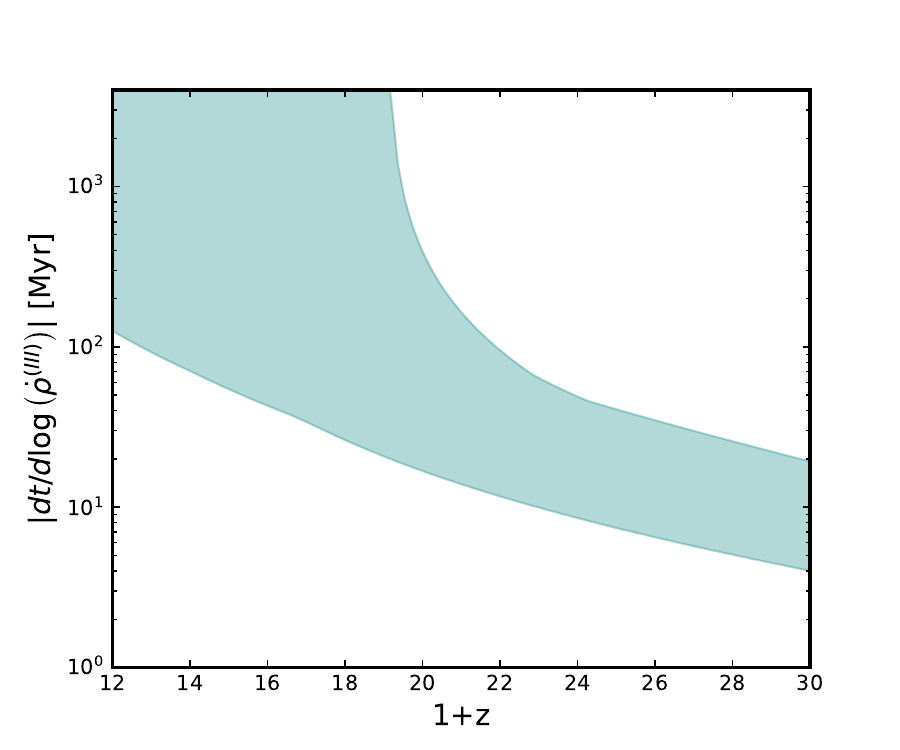}
  \caption{{\bf Left:} The emission spectra assumed here for Pop-II stars and Pop-III stars following Ref.~\cite{Barkana:2004vb} and normalized to $N_{\rm{ion}}=5000,44000$ respectively. In dashed we show the energy averaged Lyman-band Pop-II spectra, which can be used instead of the full Pop-II spectra to a good approximation. {\bf Right:} 
  The 95\% confidence envelope of the time scale for changes in the SFRD of Pop-III stars. 
  }
  \label{fig:SFRD_timescale}
\end{figure}

Since no direct observation of Pop-III stars are available, we rely on modeling to assess the validity of the approximation in Eq.~\eqref{eq:emissivityshortlife}.  For this purpose, we use the results from Ref.~\cite{Gessey_Jones_2022}, who employed the MESA stellar evolution code \cite{Paxton_2019} to simulate the evolutionary histories of individual metal-free stars of various masses, and calculated their emission spectra using the TLUSTY stellar atmosphere code \cite{hubeny1988computer}.

On the right of Fig.~\ref{fig:SFRD_timescale} we show the timescale of variations in the SFRD of Pop-III stars for all viable astrophysical models considered in our scan neglecting LW feedback. The fastest timescale sets an upper limit on stellar lifetimes at $\sim 4\,\rm{ Myr}$, beyond which the approximation in Eq.~\eqref{eq:emissivityshortlife} breaks. According to \cite{Gessey_Jones_2022} such short lifetimes are consistent with stars of mass $M \gtrsim 50 M_{\rm \odot}$. 

Pop-III stars are often assumed to have formed in isolation, leading to the expectation of a top-heavy IMF with a dominant contribution from stars with masses of order $\sim 100M_{\odot}$~\cite{Larson_1998,Bromm_2009}. In this case, the short lifetime approximation clearly remains valid. However, recent simulations challenge this assumption, and point to the possibility that Pop-III stars might have formed in clusters~\cite{Greif_2011}. In this regard~\cite{Gessey_Jones_2022} studied the emission rate of UV photons as a function of time, measured with respect to the onset of star formation, while considering different models of the IMF. They find a strong peak at $t<4{\rm Myr}$ for IMF models that motivated by simulations~\cite{Greif_2011,dopcke2013initial} and stellar archaeology studies of metal mixing~\cite{Tarumi_2020}. The only exception where the dominant contribution accumulates over a longer time period was for the Salpeter IMF, which represents an extreme case where the Pop-III IMF is primarily supported by lower-mass stars. Given these result and our current understanding of Pop-III stars we find it overall reasonable to assume the short lifetime approximation. 

\section{LW Feedback}\label{app:LW}
\begin{figure}%
\centering
\includegraphics[width=0.4\textwidth]{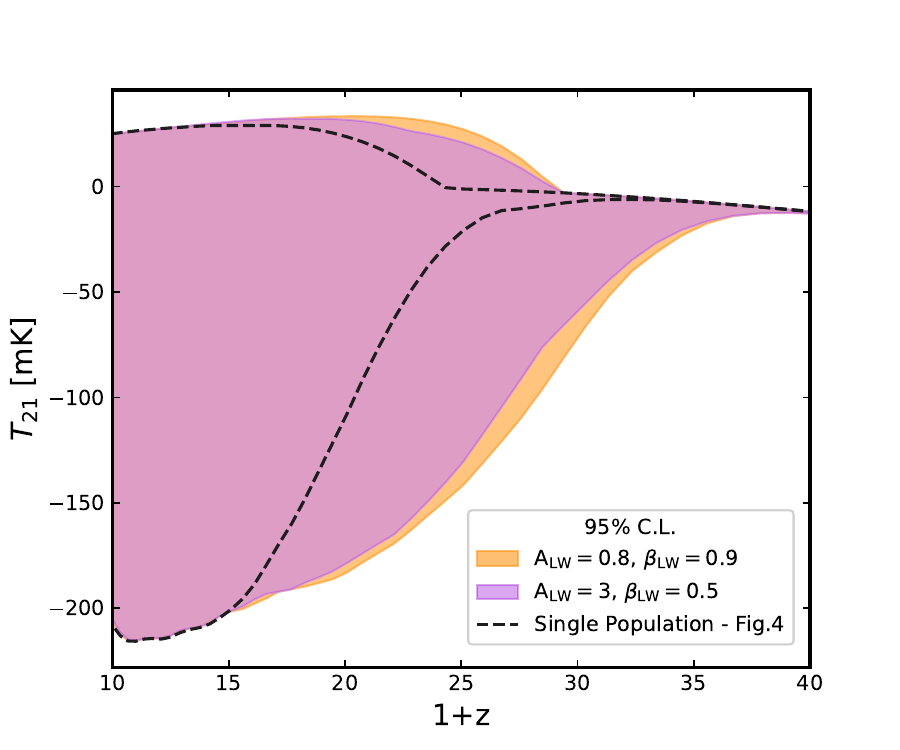}
\qquad
\caption{The viable region of 21-cm signals for three different models. The {\bf orange} ({\bf purple}) shaded regions show the 95\% confidence envelopes for the two stellar population model, taking $A_{\rm LW} = 0.8$, $\beta_{\rm LW} = 0.9$ ($A_{\rm LW} = 3$, $\beta_{\rm LW} = 0.5$), in agreement with the simulation results of~\cite{Kulkarni_2021,schauer2021influence}.
The dashed black line corresponds to the 95\% confidence envelope for the single population model, previously shown as the pink envelope in Fig.~\ref{fig:21cmsignal}. A large region of the 
$T_{21}$ parameter space exists where the two-population model can be distinguished from the single-population model for both choices of LW feedback}
\label{Fig:T21_LW_compare}
\end{figure}

\begin{figure*}%
\centering
\includegraphics[width=1\textwidth]{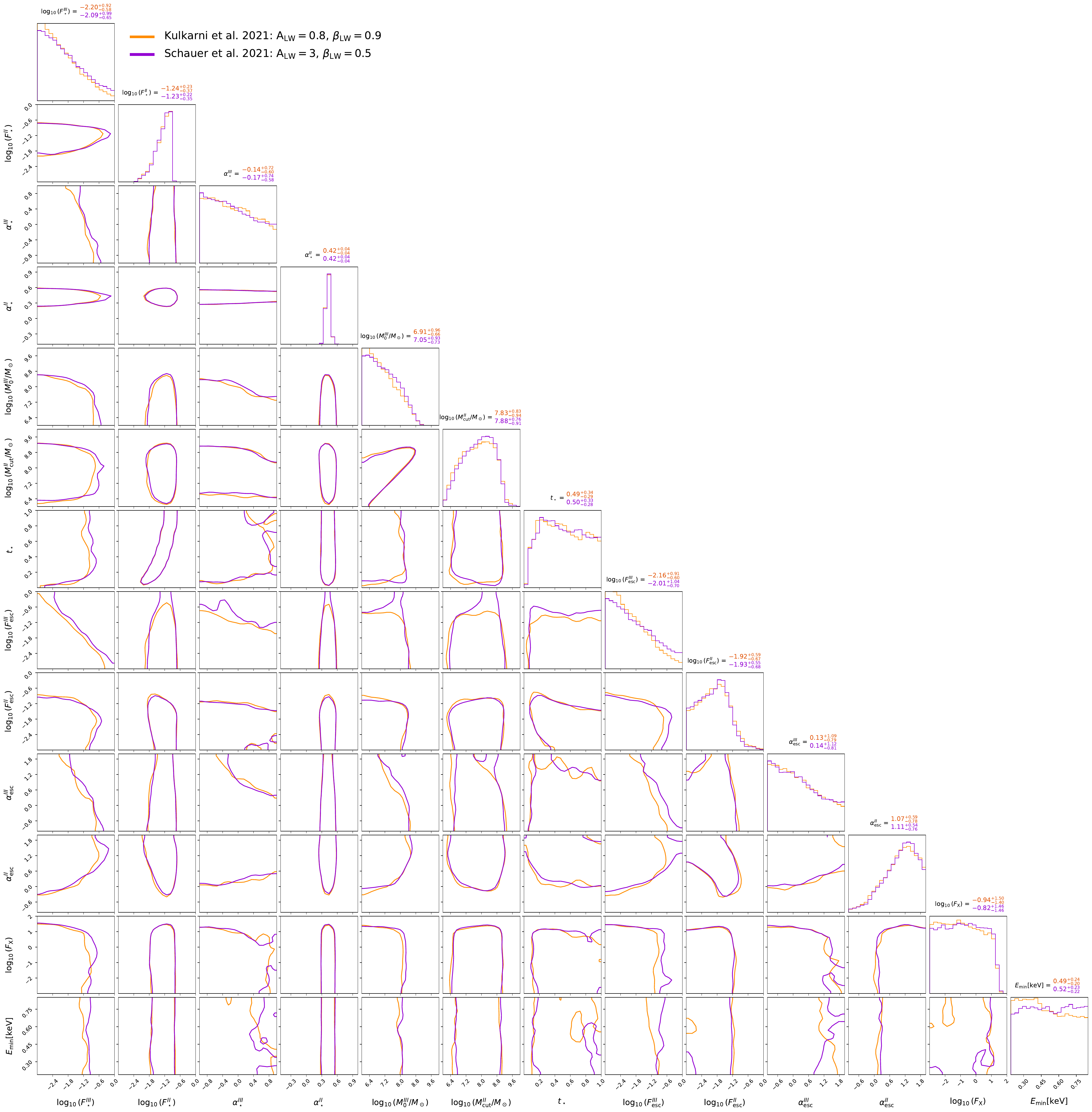}
\qquad
\caption{Corner plot of the two stellar population model (as in Fig.~\ref{Fig:MCMC}) for two different choices of LW feedback parameters~\eqref{eq:LW_feedback}: $A_{\rm LW} = 0.8$, $\beta_{\rm LW} = 0.9$ \textbf{(Orange)} and  $A_{\rm LW} = 3$, $\beta_{\rm LW} = 0.5$ \textbf{(Purple)}, corresponding to the simulation results by~\cite{Kulkarni_2021,schauer2021influence}. We find no significant difference between the two cases, or, clearly, from the central choice of $A_{\rm LW} = 2$, $\beta_{\rm LW} = 0.6$~\cite{Munoz:2021psm}, assumed in Fig.~\ref{Fig:MCMC} and throughout this paper.}
\label{Fig:MCMC_LW_compare}
\end{figure*}

Throughout this work, we adopt the LW feedback parameters from~\cite{Munoz:2021psm} (see Eq.~\eqref{eq:LW_feedback}), chosen to align with the central behavior observed in the simulations of~\cite{Kulkarni_2021,schauer2021influence}. Here we assess the sensitivity of our main results to this specific choice by reproducing them using the LW feedback parameters $(A_{\rm LW},\beta_{\rm LW}) = (0.8,0.9)$, and  $(3,0.5)$, which correspond to the findings of~\cite{Kulkarni_2021,schauer2021influence}. 
In Fig.~\ref{Fig:MCMC_LW_compare}, we show the corner plots for both LW parameter sets, which we produced following the procedure described in Sec.~\ref{sec:constraints} and App.~\ref{sec:Likelihood}.
We observe only a slight variations in the distributions between the two scenarios. Fig.~\ref{Fig:T21_LW_compare} shows the corresponding 21-cm global signal 95\% confidence envelopes, along with that of the single-population (previously shown as the pink region in Fig.~\ref{fig:21cmsignal}). While the choice of 
$(A_{\rm LW},\beta_{\rm LW}) = (3,0.5)$ results in a slightly smaller envelope, it remains significantly larger than that of a single population, still leaving a broad region of the $T_{21}$ parameter space where the presence of a second population can be identified.

For completeness, we write the equation used to calculate the LW flux produced by a single stellar population
\begin{equation}
    J_{\rm LW}^i = \frac{(1+z)^2}{4\pi} \frac{\bar{\eps}^i_{LW}}{\mu_b} \int \frac{\dot{\rho}^i(z')}{H(z')}f_{\rm LW}\left(\frac{1+z'}{1+z}\right)
\end{equation}
$\bar{\eps}^i_{LW}$ is the Pop-i Lyman band emissivity (Eq.\eqref{eq:eL}) averaged over the LW energies, and $f_{\rm LW}\left(\frac{1+z'}{1+z}\right)$ accounts for the attenuation due to absorption by hydrogen atoms, and is taken from~\cite{Fialkov_2013} (also see~\cite{ahn2009inhomogeneous,qin2020tale}).

\end{document}